\documentclass[]{aastex631}
\usepackage{amsthm}
\usepackage{amsmath}
\usepackage{float} 
\usepackage{bm} % For math bold fonts.

\received{November 06, 2025}
\revised{February 01, 2026}
\accepted{February 03, 2026}
\submitjournal{ApJ [Accepted]}

\shorttitle{KH Instability with Pressure Anisotropy}
\shortauthors{Biswas et al.}
\begin{document}

\title{Two-Dimensional Kelvin-Helmholtz Instability with Anisotropic Pressure}

\correspondingauthor{Dinshaw S. Balsara}
\email{dbalsara@nd.edu}

\author[0000-0002-4879-8889]{Shishir Biswas}
\affiliation{Department of Physics and Astronomy, University of Notre Dame, Notre Dame, IN, USA}

\author[0000-0002-7203-0730]{Masaru Nakanotani}
\affiliation{Department of Space Science, University of Alabama in Huntsville, Huntsville, AL, USA}
\affiliation{Center for Space Plasma and Aeronomic Research, University of Alabama in Huntsville, Huntsville, AL, USA}

\author[0000-0003-3309-1052]{Dinshaw S. Balsara}
\affiliation{Department of Physics and Astronomy, University of Notre Dame, Notre Dame, IN, USA}
\affiliation{ACMS Department, University of Notre Dame, Notre Dame, IN, USA}

\author[0000-0001-5485-2872]{Vladimir Florinski}
\affiliation{Department of Space Science, University of Alabama in Huntsville, Huntsville, AL, USA}
\affiliation{Center for Space Plasma and Aeronomic Research, University of Alabama in Huntsville, Huntsville, AL, USA}

\author[0000-0002-8767-8273]{Merav Opher}
\affiliation{Astronomy Department, Boston University, Boston, MA 02215, USA}

\begin{abstract}
The Kelvin-Helmholtz (KH) instability occurs in multiple heliospheric (solar-wind stream interfaces, planetary magnetospheres, cometary tails, heliopause flanks) and interstellar (protoplanetary disks, relativistic jets, neutron star accretion disks) environments. While the KH instability has been well-studied in the magnetohydrodynamic (MHD) limit, only limited studies were performed in the collisionless regime, which is conducive to development of anisotropic pressures. Collisionless plasmas are often described using the Chew Goldberger and Low (CGL) equations which feature an anisotropic pressure tensor. This paper presents a comprehensive analysis of the CGL version of the KH instability using linearised and numerical techniques. We find that the largest growth rates and the greatest incidence of magnetic effects occur in the MHD limit. In the large relaxation time CGL limit, part of the energy goes into the formation of pressure anisotropies, resulting in smaller amounts of energy being available for bending the field lines. Consequently, when we cross-compare CGL and MHD simulations that are otherwise identical, the current densities are largest in the MHD limit, and the largest magnetic islands also form in that limit. Early and late time formation of pressure anisotropies have also been studied. We also find that the strongest trend for forming intermittencies in the flow also occurs in the MHD limit. The paper also discusses possible consequences of our results for turbulence and reconnection in the heliosheath (the layer between the solar wind termination shock and the heliopause).
\end{abstract}

\keywords{Astrophysical fluid dynamics, CGL, KH Instability, Magnetic Reconnection, Magnetic Islands }

%%%%%%%% Section:1 Introduction %%%%%%%%

\section{Introduction} \label{sec:introduction}
Shear-driven instabilities, such as the Kelvin–Helmholtz (KH) instability, are ubiquitous in astrophysical and space plasmas, arising wherever velocity gradients or shear layers are present. The KH instability plays a crucial role in the dynamics of astrophysical jets \citep{Baty_2002, Viallet_2007, Nishikawa_2014}, where significant velocity shear arises between the rapidly out-flowing jet plasma and the surrounding medium. The development of KH modes at the jet–ambient interface can result in the absorption and mixing of external material into the jet, therefore modifying its density contrast, momentum transport, and large-scale configuration. These instabilities also induce turbulence and vortical structures, which promote energy dissipation, magnetic field amplification, and particle acceleration, processes intimately associated with the observed non-thermal emission from jets. Analogous phenomena are relevant in solar chromospheric \citep{Kuridze_2016} and coronal jets \citep{Zhao_2018}, where the interaction between high-velocity outflows and the surrounding plasma can induce KH waves, facilitating heating, mixing, and structuring of the solar environment.

In accretion disks, velocity shear naturally appears both radially and vertically. These shear layers provide favorable conditions for the development of KH instabilities, which can result in turbulence, angular momentum transport, and enhanced mixing between various disk regions \citep{Balbus_1998, Lovelace_2010}. Since the KH instability is a velocity shear driven instability, as is the magnetorotational instability (MRI), a study of KH instability may generate insights that could also be relevant for the MRI.

Although KH instability is a central feature of numerous astrophysical systems, it is also of essential importance in plasma physics. The KH instability in plasma physics illustrates how shear flows in both magnetized and unmagnetized plasmas can generate coherent vortical structures and turbulence \citep{Frank_APJ_1996, Jones_1997, Miura_PoP_1997, KEPPENS_1999, Ryu_2000, Jeong_2000, Li_SciRep_2023}. Laboratory plasma experiments have revealed the emergence of KH modes in regulated shear setups, providing significant understanding of nonlinear saturation, momentum transport, and energy dissipation. The coexistence of magnetic reconnection signatures with KH instability has also been shown to strongly modify the system dynamics, as revealed in both fluid simulations \citep{KEPPENS_1999, nakamura2006, Li_SciRep_2023} and particle-in-cell (PIC) simulations \citep{nakamura2017,  Ahmadi_PoP_2025}. These studies not only enhance our comprehension of fundamental plasma processes, but also function as analogies for the interpretation of the behavior of astrophysical shear flows. Thus, KH instability arises as a unified concept pertinent to both laboratory plasmas and many astrophysical contexts.

The KH instability is crucial in space plasmas, especially at interfaces where rapidly moving plasma interacts with a slower medium. Notable situations encompass the Earth's magnetopause \citep{ONG_1972, Hasegawa_2004, Quijia_2021}, planetary magnetotails \citep{Otto_2000}, the boundaries between the solar wind and planetary atmospheres \citep{Johnson_2014}, and the heliopause where the solar wind engages with the interstellar medium \citep{Borovikov_Pogorelov_Zank_Kryukov_2008, Florinski_2025}.  KH instabilities are also detected in coronal mass ejections (CMEs) \citep{Mostl_2013}, where significant velocity shear at the CME-solar wind interface generates ripples and vortical structures that affect CME propagation, magnetic topology, and space weather impacts on Earth. Wave signatures of KH instabilities were found in shear layers in the vicinity of the heliospheric current sheet \citep{Kieokaew_Lavraud_Yang_etal_2021}. Satellite observations have also captured KH-like ripples along the flanks of CMEs. Velocity shear within the plasma depletion layer (PDL), which develops upstream of planetary magnetopauses \citep{Zwan_1976}, can also induce KH instabilities.

PDLs can also develop in the heliosphere at shear boundaries, such as in regions where the rapid solar wind interacts with slower plasma flows or near the heliopause \citep{Cairns_2017, Florinski_2025}.  Velocity shear at the boundaries of these PDLs can induce KH instabilities, producing vortices and turbulence that facilitate plasma mixing and energy transfer on heliospheric scales. The potential for KH instability induced by velocity shear at the heliopause has been addressed in previous studies by \citet{Chalov_1996} and \citet{Avinash_2015}. These instabilities are also crucial in the context of heliospheric jets, as reported by \citet{Ma_2024}. The outer heliosphere has been confirmed to have regions of reduced plasma density and an enhanced magnetic field by observations from Voyager 1 \& Voyager 2 \citep{Gurnett_Kurth_2019}. This is consistent with the formation of PDLs and supports the presence of KH-like activities at heliospheric boundaries.  These processes illustrate that, akin to planetary magnetospheres and coronal mass ejections, KH instability significantly influences the formation and dynamics of PDL structures on heliospheric scales. Hence, grasping the emergence of these instabilities in the heliosheath is essential for understanding the dynamic processes within the heliosphere and its large-scale structure.

Although the aforementioned observations and discussions are primarily analyzed within the magnetohydrodynamic (MHD) paradigm, the majority of astrophysical and space plasmas are, in fact, dilute and exhibit weak collisionality. In such dilute plasmas, the Coulomb collisions are infrequent with the result that the velocities of the particles that make up the plasma may deviate from a Maxwell-Boltzmann distribution. As pointed out by Chew, Goldberger and Low (\citet{Chew_Goldberger_Low_1956}, CGL henceforth), in such situations, the system can exhibit two distinct pressures, one parallel and the other perpendicular to the magnetic field, resulting in a pressure anisotropy. Both the MHD and CGL equations are derived from a fluid approximation and Faraday's law, but the CGL model incorporates the double-adiabatic approximation, which is particularly relevant for dilute plasmas. Importantly, CGL and MHD are different both theoretically and numerically, as described by \citet{kato_1966} and \citet{Bhoriya_Balsara_Florinski_Kumar_2024}. The CGL equations support a different set of waves compared to the MHD equations. The pressure anisotropy in CGL system substantially modifies the dynamics of shear flows, instabilities, and plasma structures, potentially altering the growth and saturation of KH modes, impacting turbulence, and effecting energy and momentum transmission. Therefore, to comprehensively understand KH instabilities and related phenomena in dilute plasmas - encompassing astrophysical jets, accretion disks, magnetospheric secnerios and heliospheric boundaries - it is crucial to investigate models that go beyond the MHD approximation; with its assumption of an isotropic pressure. 
The CGL model \citep{Chew_Goldberger_Low_1956} offers a natural extension of magnetohydrodynamics (MHD) to characterize collisionless or weakly collisional plasmas exhibiting pressure anisotropy. The CGL framework accounts for the influence of anisotropy on plasma dynamics by independently evolving the pressures parallel and perpendicular to the magnetic field. The CGL theory anticipates two instabilities, firehose and mirror, associated to the Alfv\'{e}n and slow magnetosonic waves, respectively, as posited by \citet{Parker_1958, Chandrasekhar_1958}, and \citet{kato_1966}.
The CGL model facilitates the investigation of how anisotropic pressures affect the emergence, growth rate, and nonlinear evolution of KH phenomena, along with the consequent turbulence. In recent years, several analytical efforts have been undertaken to comprehend the KH instability in anisotropic solar wind plasma \citep{Ismayilli_2018} and in anisotropic space plasma \citep{Dzhalilov_MNRAS_2023}. However, numerical efforts to comprehend its dynamics in both the linear and nonlinear regimes remain inadequate, largely due to the associated numerical challenges.

For a long time, the CGL equations were regarded as numerically intractable. This has changed with the paper of \citet{Bhoriya_Balsara_Florinski_Kumar_2024} who found ways to overcome all the numerical challenges inherent in the CGL equations. An initial application of these ideas, as they pertain to the heliosphere, was also made \citep{Florinski_2025}. The goal of this paper is to investigate the KH instability within the context of the CGL model. We seek to quantify the influence of pressure anisotropy on the linear growth and nonlinear saturation of KH instability by directly comparing the CGL results with those derived from the usual MHD limit. Our simulations enable us to investigate how anisotropy alters the outcome of KH instability. Additionally, we perform a direct comparison between the KH growth rates derived from our numerical simulations and the linearised predictions for various CGL relaxation times, thereby establishing a direct connection between theory and computations. This study offers novel insights into the dynamics of shear flows in anisotropic space and astrophysical plasmas, connecting linearised models with the numerical understanding.

This paper is organized as follows. In Section \ref{sec:model}, we describe the set of governing equations and our simulation setup. Section \ref{sec:linertheory} discusses the linear theory of the CGL KH instability, and in Section \ref{sec:results}, we present the visual data analysis. Section \ref{sec:diagnostics} is dedicated to various diagnostics that we have performed. Finally, we conclude with a summary and future directions in Section \ref{sec:conclusion}. 

%%%%%%%% Section:2 Equations & Model %%%%%%%%

\section{Governing Equations \& Simulation Setup} \label{sec:model}

\subsection{Governing Equations}
In a dense, collisional plasma, frequent particle collisions drive the velocity distribution toward a Maxwell-Boltzmann distribution, resulting in an isotropic pressure that can be treated like a single scalar. In contrast, most astrophysical plasmas are dilute and weakly collisional, allowing them to develop pressure anisotropy, where the pressure parallel and perpendicular to the magnetic field can differ. This leads to pressure anisotropy, which is described by the CGL theory \citep{Chew_Goldberger_Low_1956}.
In contrast to ideal (isotropic) MHD, the pressure tensor in the CGL system \citep{Chew_Goldberger_Low_1956} consists of two distinct components, which are orthogonal ($p_\perp$) and parallel ($p_\parallel$) to the magnetic field $\mathbf{B}$. In a stationary inertial reference frame, the pressure tensor is represented as
\begin{equation}
\label{eqn_prstens}
\mathbf{p}=p_\perp\mathbf{I}+(p_\parallel-p_\perp)\mathbf{bb},
\end{equation}
where $\mathbf{b}=\mathbf{B}/B$ denotes the unit vector aligned with the magnetic field direction. Due to pressure anisotropy in collisionless or weakly collisional plasmas, the plasma pressure parallel to and perpendicular to the magnetic field can differ. As a result, two separate plasma $\beta$ parameters are defined:
\[
\beta_\parallel = \frac{8 \pi p_\parallel}{B^2}, \qquad
\beta_\perp = \frac{8 \pi p_\perp}{B^2},
\]
which quantify the relative importance of plasma pressure to magnetic pressure in directions that are parallel to and perpendicular to the magnetic field \citep{Parker_1958, Chandrasekhar_1958}. \citet{Chandrasekhar_1958} and \citet{kato_1966} further noted that the degree of pressure anisotropy in a plasma is limited by the onset of kinetic instabilities. In particular, the firehose  instability occurs when the parallel pressure exceeds the perpendicular pressure by a substantial margin  (\(p_\parallel > p_\perp \)), while the mirror instability arises when the perpendicular pressure dominates by a substantial margin (\(p_\perp > p_\parallel\)), with the precise instability thresholds depending on the plasma $\beta$. Most importantly, the CGL system can lose hyperbolicity when the pressure anisotropy ratio  ($A = {p_\perp}/{p_\parallel}$) becomes significantly larger or smaller than unity, as discussed by \citet{kato_1966} and more recently by \cite{Bhoriya_Balsara_Florinski_Kumar_2024}.

The resulting system comprises five fundamental physical variables-mass density ($\rho$), fluid velocity ($\mathbf{v}$), parallel pressure ($p_\parallel$), perpendicular pressure ($p_\perp$), and magnetic field ($\mathbf{B}$). The governing equations form a $9\times9$ hyperbolic system. Following \citet{Bhoriya_Balsara_Florinski_Kumar_2024}, the CGL equations can be formulated in cgs units as a non-conservative hyperbolic system accompanied by a stiff source term, as outlined below:

% ========================
\begin{align} 
% ========================
% Mass
\frac{ \partial \rho}{\partial t}+\nabla \cdot (\rho \mathbf{v})&=0
\label{eq:mass_conv} \\
% ========================
% Momentum
\frac{\partial (\rho \mathbf{v})}{\partial t}
+
\nabla \cdot 
\left[\rho \mathbf{v} \mathbf{v}
+p_\perp \textbf{I}+(p_\parallel-p_\perp) \mathbf{ b} \mathbf{ b} -\frac{1}{4\pi}
\left(\mathbf{B}\mathbf{B}-\frac{B^2}{2}\textbf{I}\right) \right]
&=
0
\label{eq:momentum_conv} \\
% ========================
% Pressure jump
\frac{\partial (p_\parallel-p_\perp)}{\partial t}
+
\nabla \cdot \left[(p_\parallel-p_\perp)\mathbf{v}\right] +(2p_\parallel+p_\perp )\times\mathbf{ b}\cdot\nabla \mathbf{v}\cdot\mathbf{ b} - p_\perp\nabla\cdot\mathbf{v}
&=
-\frac{1}{\tau} \left( p_\parallel - p_\perp \right)
\label{eq:press_jump} \\
% ========================
% Energy
\frac{\partial \mathcal{E}}{\partial t} +
\nabla \cdot \left[\mathbf{v}\left(\mathcal{E} + p_\perp+ \frac{B^2}{8 \pi}\right)
+
\mathbf{v} \cdot \left((p_\parallel-p_\perp) \mathbf{ b} \mathbf{ b} -\frac{\mathbf{B} \mathbf{B}}{4 \pi}\right)\right]&=0
\label{eq:energy_conv}\\
% ========================
% Induction
\frac{\partial \mathbf{B}}{\partial t}+\nabla \times \left[-(\mathbf{v} \times \mathbf{B})\right]&=0
\label{eq:induction}
% ========================
\end{align}
% ========================

The system is closed using the following equation of state:
\begin{equation}
    \mathcal{E}=\dfrac{1}{2}\rho {v}^2 + \frac{B^2}{8 \pi} + \frac{3}{2} \bar{p}
    \label{eq:EOS}
\end{equation}
where $\bar{p}\equiv \dfrac{2 p_\perp + p_\parallel}{3}$ is the average pressure. In the aforementioned set of equations, the relaxation time is represented by $\tau$. This relaxation time can be set by microphysical processes in a small scale simulation or it can be set by pitch angle scattering with respect to ambient turbulence \citep{Jokipii_1966} in a simulation that operates on macrophysical timescales, as shown in \citet{Florinski_2025}. Since this is primarily a parameter study to understand the role of relaxation in the CGL equations, as applied to the Kelvin Helmholtz (KH) instability, we will explore simulations with different values of $\tau$ that are preset. This is also more meaningful for this paper because one of the goals of this work is to develop linear theory for the KH instability in the CGL context for the very first time. Our goal is to then show that the growth rates can be matched with numerical simulations, thereby providing a strong validation of our numerics.

To solve the CGL equations described above, we use a third-order accurate Godunov scheme ensuring a divergence-free evolution of the magnetic field. Details of the numerical methods are provided in \citet{Bhoriya_Balsara_Florinski_Kumar_2024, Balsara_CAMC_2025, Balsara_APJ_2025}.

\subsection{Simulation Setup}
In studies of KH instability, using an abrupt (discontinuous) profile  of velocity, density, and magnetic field leads to instability at the smallest scales that can be represented in a code; i.e. we would get fastest growing instabilities on the scale of one or a few zone sizes. As a result, the growth of these small-scale modes would dominate, making it difficult  to compare the linearised and numerical growth rates of the instability. To avoid this issue, we adopt a smooth $\tanh$ profile for the velocity, density, and magnetic field, which suppresses the dominance of the smallest scales and allows a more meaningful comparison between analytical predictions and numerical results.

\begin{figure}%[H]
	\begin{center}
		\includegraphics[width=0.55\textwidth,clip=]{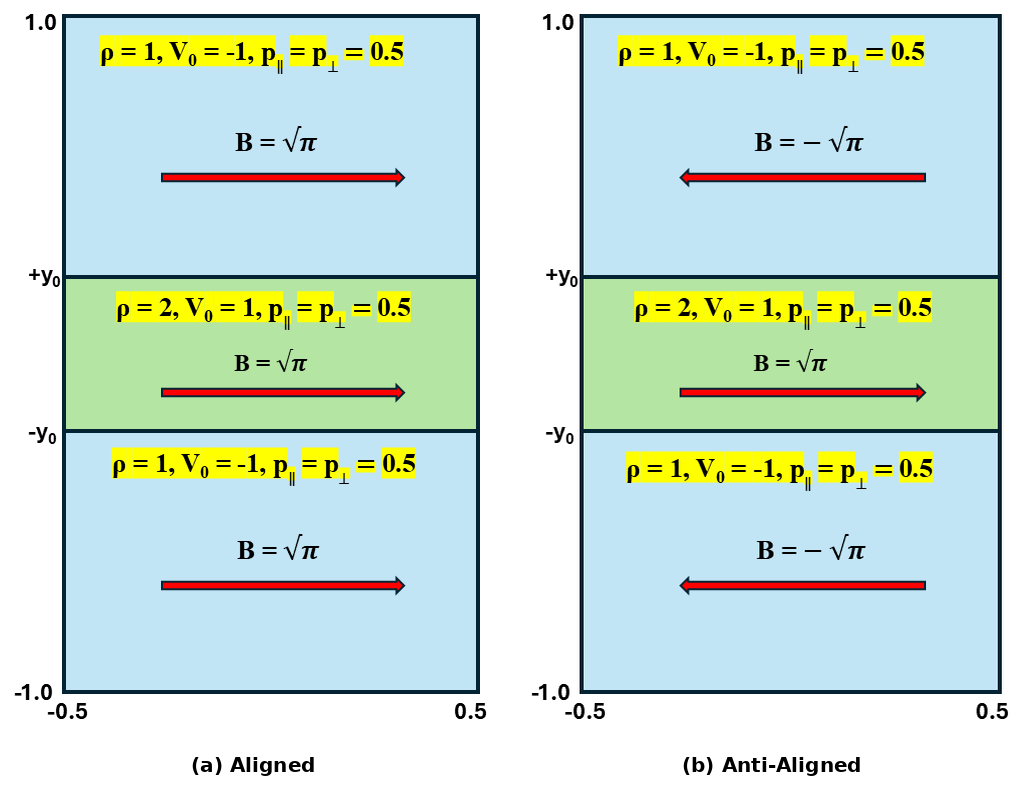}
		\caption{Simulation setup for the 2D KH instability, showing the domain, initial velocity, density, pressure, magnetic field configuration, and flow directions. Panel (a) illustrates the aligned magnetic field configuration, with magnetic field vectors pointing to the right throughout the domain, while panel (b) shows the anti-aligned configuration, where the magnetic field vectors point to the right in the central region and to the left in the upper and lower regions.}
		\label{fig_1}
	\end{center}
\end{figure}

Based on the above governing equations, we set up the initial equilibrium state for the KH instability in the $xy$ plane. The initial profiles of density, $x$-velocity, and magnetic field are prescribed as follows to establish the shear layer and magnetic configuration:

% ========================
% Initial Conditions
% ========================
\begin{align}
	\rho &= 1 + \frac{1}{2} 
	\Bigg[
	\tanh\Big(\frac{y_0 - y}{h}\Big) 
	+ \tanh\Big(\frac{y + y_0}{h}\Big)
	\Bigg] 
	\label{eq:rho}
\end{align}

\begin{align}
	v_x &= v_0 
	\Bigg[
	\tanh\Big(\frac{y_0 - y}{h}\Big) 
	+ \tanh\Big(\frac{y + y_0}{h}\Big) 
	- 1
	\Bigg] 
	\label{eq:vx}
\end{align}

\begin{align}
	B_x &= B_0 
	\Bigg[
	\tanh\Big(\frac{y_0 - y}{h}\Big) 
	+ \tanh\Big(\frac{y + y_0}{h}\Big) 
	- 1
	\Bigg] 
	\label{eq:Bx}
\end{align}
Here, $y_0 = 0.25$ specifies the half-separation between the two shear interfaces, $h = 0.03$ is the characteristic thickness of each shear layer, $v_0 = 1.0$ is the shear flow velocity amplitude, and $B_0 = \sqrt{\pi}$ is the initial magnetic field strength. This configuration represents two symmetric, anti-parallel shear layers about $y=0$, as illustrated schematically in Figure \ref{fig_1}.

The initial plasma pressure is assumed to be uniform, with both parallel and perpendicular components set to $p_\parallel = p_\perp = 0.5$.  The corresponding plasma $\beta$, defined as the ratio of gas pressure to magnetic pressure, is 
\[
\beta_\parallel = \beta_\perp = \frac{0.5}{B_0^2 / 8\pi} = 4.0.
\]

% ========================
% Velocity Perturbations
% ========================
To trigger the KH instability, a small transverse velocity perturbation is introduced in the $y$-direction:

\begin{align}
	v_y &= \epsilon \sin(2\pi x) 
	\left\lbrace\exp\left[-\frac{(y - y_0)^2}{h^2}\right] 
	+ \exp\left[-\frac{(y + y_0)^2}{h^2}\right]\right\rbrace,
	\label{eq:vy}
\end{align}
where $\epsilon = 0.02$ denotes the amplitude of the perturbation.

% =============================
% Figure 1: KH Simulation Model
% =============================

With the aforementioned initial conditions, the governing system of equations \eqref{eq:mass_conv}--\eqref{eq:induction} is evolved using a third-order finite-volume WENO scheme as described in \citet{Bhoriya_Balsara_Florinski_Kumar_2024}. The computational domain spans $[-0.5,0.5]$ in the x-direction and $[-1.0,1.0]$ in the y-direction, discretized with a uniform grid of $600\times1200$ cells. We have also carried out a grid-resolution convergence test by reducing the spatial resolution to 75\% (i.e., $450 \times 900 $) of the original grid in each direction. We find that the time evolution and growth rate of the magnetic energy are indistinguishable between the original ($600 \times 1200$) and the reduced-resolution ($450 \times 900$) runs, demonstrating that the results presented are numerically converged with respect to grid resolution. The grid-resolution convergence results are illustrated in Fig. \ref{fig_6} (Section \ref{subsec:growth rate}), with a detailed discussion provided. Periodic boundary conditions were enforced in both directions. For our shear flow simulation setup, the characteristic eddy turnover time is estimated as 
\[
\tau_{\rm eddy} \sim \frac{L}{v_0} \approx 0.5,
\] 
where $L=2y_0$ is the thickness of the shear flow channel and $v_0$ is the shear velocity. Simulations were carried out up to 24 eddy turnover times ($\tau_{\rm eddy}$), by which time the numerical data indicated that the system has reached a fully nonlinear state.

As illustrated in Figure \ref{fig_1}, our simulation model includes both aligned and anti-aligned magnetic field configurations. It is noteworthy that, inside the heliosheath, different aligned and anti-aligned magnetic field configurations may occur depending on the location. Global evidence of such aligned and anti-aligned magnetic field configurations has recently been reported by \cite{Ma_2025}. To account for this natural variability, we utilize both types of configurations as initial conditions in our simulations of the KH instability. The comprehensive list of simulations used in this study is summarized in Table \ref{table_1_Simulation_Runs}.

% =============================
% Table 1: Simulation Runs
% =============================
\begin{table}%[H]
\begin{center}
\caption{Overview of the simulation runs presented in this study, including the initial magnetic field configuration and the CGL relaxation times employed to investigate the evolution of the KH instability.}
\label{table_1_Simulation_Runs}
\begin{tabular}{cccccccccc}
\tableline
  Runs  & $\tau$    &   $\tau / \tau_{\rm eddy}$  &   Magnetic field orientation \\
\tableline
Run1   & $5$        &  $10$                       & Anti-Aligned \\
Run2   & $1$        &  $2$                        & Anti-Aligned \\
Run3   & $0.1$      &  $0.2$                      & Anti-Aligned \\
Run4   & $10^{-5}$  &  $2 \times 10^{-5}$         &  Anti-Aligned \\
Run5   & $5$        & $10$                        &  Aligned \\
Run6   & $1$        &  $2$                        & Aligned \\
Run7   & $0.1$      &  $0.2$                      & Aligned \\
Run8   & $10^{-5}$  &  $2 \times 10^{-5}$         & Aligned \\
\tableline
\end{tabular}
\end{center}
\end{table}

As the shear layer width is set to $h = 0.03$, while the grid resolution in the $y$-direction is 
$\Delta y = 2/1200$, this gives
\[
\frac{h}{\Delta y} = \frac{0.03}{2/1200} = 18,
\]
This means that the scale length of the shear layer is resolved over 18 grid points. With this resolution, 
we ensure that the system (in its initial evolution) does not develop very small scale numerically induced instabilities on the scale of one or a few mesh zones. This enables us to make a reliable comparison between the numerical growth rates and the linearised predictions  of the KH instability. 
In the following section, we sketch the linearised calculation of the KH instability growth rates in the CGL limit.

%%%%%%%% Section:3 Linear Theory %%%%%%%%

\section{Discussion of linear theory} \label{sec:linertheory}
In this section, we explain the linear analysis of the CGL system, which solves the governing equations as an initial value problem.
The same method has been used in \citet{miura1982} and \citet{fujimoto1991}.
We use the following anisotropic pressure equations instead of (\ref{eq:press_jump}) and (\ref{eq:energy_conv}),
\begin{gather}
\frac{D}{Dt}\left(\frac{p_\perp}{\rho B}\right)=0; \\
\frac{D}{Dt}\left(\frac{p_\parallel B^2}{\rho^3}\right)=0.
\end{gather}
Linearizing and applying a Fourier transformation for the governing equations by $\Psi=\Psi_0(y)+\Psi_1(t,y)e^{ik_xx}$ for $\rho$, $v_x$, $p_\perp$, $p_\parallel$, $B_x$, and $B_z$ and $\Phi=\Phi_1(t,y)e^{ik_xx}$ for $v_y$, $v_z$, and $B_y$, we obtain the following linearized equations,
\begin{align}
&\partial_t\rho_1+v_{x0}ik_x\rho_1+v_{y1}\partial_y\rho_0=-\rho_0(\widetilde{\nabla\cdot{\bf v}})_1; \\
&\begin{aligned}
\rho_0\partial_tv_{x1}&+\rho_0v_{x0}ik_xv_{x1}+\rho_0v_{y1}\partial_yv_{x0}\\
&+ik_x(B_{x0}B_{x1}+B_{z0}B_{z1})-B_{x0}ik_xB_{x1}-B_{y1}\partial_yB_{x0}\\
&+ik_xp_{\perp 1}+ik_x(p_{\parallel 1}-p_{\perp 1})b_{x0}b_{x0}+ik_x(p_{\parallel 0}-p_{\perp 0})2b_{x1}b_{x0}+\partial_y\{(p_{\parallel 0}-p_{\perp 0})b_{y1}b_{x0}\}=0;
\end{aligned}\\
&\begin{aligned}
\rho_0\partial_tv_{y1}&+\rho_0v_{x0}ik_xv_{y1}\\
&+\partial_y(B_{x0}B_{x1}+B_{z0}B_{z1})-B_{x0}ik_xB_{y1}\\
&+\partial_yp_{\perp 1}+ik_x\{(p_{\parallel 0}-p_{\perp 0})b_{x0}b_{y1}\}=0;
\end{aligned}\\
&\begin{aligned}
\rho_0\partial_tv_{z1}&+\rho_0v_{x0}ik_xv_{z1}\\
&-B_{x0}ik_xB_{z1}-B_{y1}\partial_yB_{z0}\\
&+\partial_y\{(p_{\parallel 0}-p_{\perp 0})b_{y1}b_{z0}\}+ik_x\{(p_{\parallel 1}-p_{\perp 1}b_{x0}b_{z0}+(p_{\parallel 0}-p_{\perp 0}b_{x0}b_{z1}\}=0;
&\end{aligned}\\
&\partial_tp_{\perp 1}+v_{x0}ik_xp_{\perp 1}=p_{\perp 0}[\widetilde{{\bf b}\cdot({\bf b}\cdot\nabla{\bf v})}]_1-2p_{\perp 0}(\widetilde{\nabla\cdot{\bf v}})_1+\frac{1}{3\tau}(p_{\parallel1}-p_{\perp1});\\
&\partial_tp_{\parallel 1}+v_{x0}ik_xp_{\parallel 1}=-2p_{\parallel 0}[\widetilde{{\bf b}\cdot({\bf b}\cdot\nabla{\bf v})}]_1-p_{\parallel 0}(\widetilde{\nabla\cdot{\bf v}})_1-\frac{2}{3\tau}(p_{\parallel1}-p_{\perp1});\\
&\partial_tB_{x1}+B_{x0}(\widetilde{\nabla\cdot{\bf v}})_1+v_{y1}\partial_yB_{x0}+v_{x0}ik_xB_{x1}-B_{y1}\partial_yv_{x0}-B_{x0}ik_xv_{x1}=0;\\
&\partial_tB_{y1}+v_{x0}ik_xB_{y1}-B_{x0}ik_xv_{y1}=0;\\
&\partial_tB_{z1}+B_{z0}(\widetilde{\nabla\cdot{\bf v}})_1+v_{y1}\partial_yB_{z0}+v_{x0}ik_xB_{z1}-B_{x0}ik_xv_{z1}=0.
\end{align}
Here, ${\bf b}_{0,1}={\bf B}_{0,1}/B_0$, and
\begin{gather}
(\widetilde{\nabla\cdot{\bf v}})_1=ik_xv_{x1}+\partial_yv_{y1};\\
[\widetilde{{\bf b}\cdot({\bf b}\cdot\nabla{\bf v})}]_1=b_{x0}^2ik_xv_{x1}+b_{x0}b_{y1}\partial_yv_{x0}+b_{x0}b_{z0}ik_xv_{z1}.
\end{gather}
Since we compare the linear analysis with 2D CGL simulations, we have neglected the wavevector along the z-direction, namely $k_z=0$.

We numerically solve the linearized equations as an initial value problem to obtain the linear growth rate and eigenfunctions. We use the same background plasma and magnetic field profiles as the nonlinear simulation.
Therefore, the boundary conditions are periodic.
We set $1000$ grid points along the y-direction.
We use the 4th-order Runge-Kutta scheme for time integration and the spectral method to evaluate the spatial derivative $\partial_y$. 
The simulation is evolved for a long enough time that the fastest growing mode dominates other modes, and then, to obtain the growth rate, we calculate the time evolution of $<B_{y1}^2>$, where $<\cdot>$ is the spatial average over the y-axis, and fit by the exponential function $\exp(2\Gamma t)$ to find the growth rate $\Gamma$. The findings from the linearised study will be presented in Section \ref{subsec:growth rate}. In that Section, we will compare the growth rates of the KH instability obtained from linear analysis with those measured from the early, linear phase of our fully nonlinear simulations.

%%%%%%%% Section:4 Visual data and Discussion %%%%%%%%

\section{Visual data analysis and discussion} \label{sec:results}
\begin{figure}%[H]
	\begin{center}
		\includegraphics[width=1.0\textwidth,clip=]{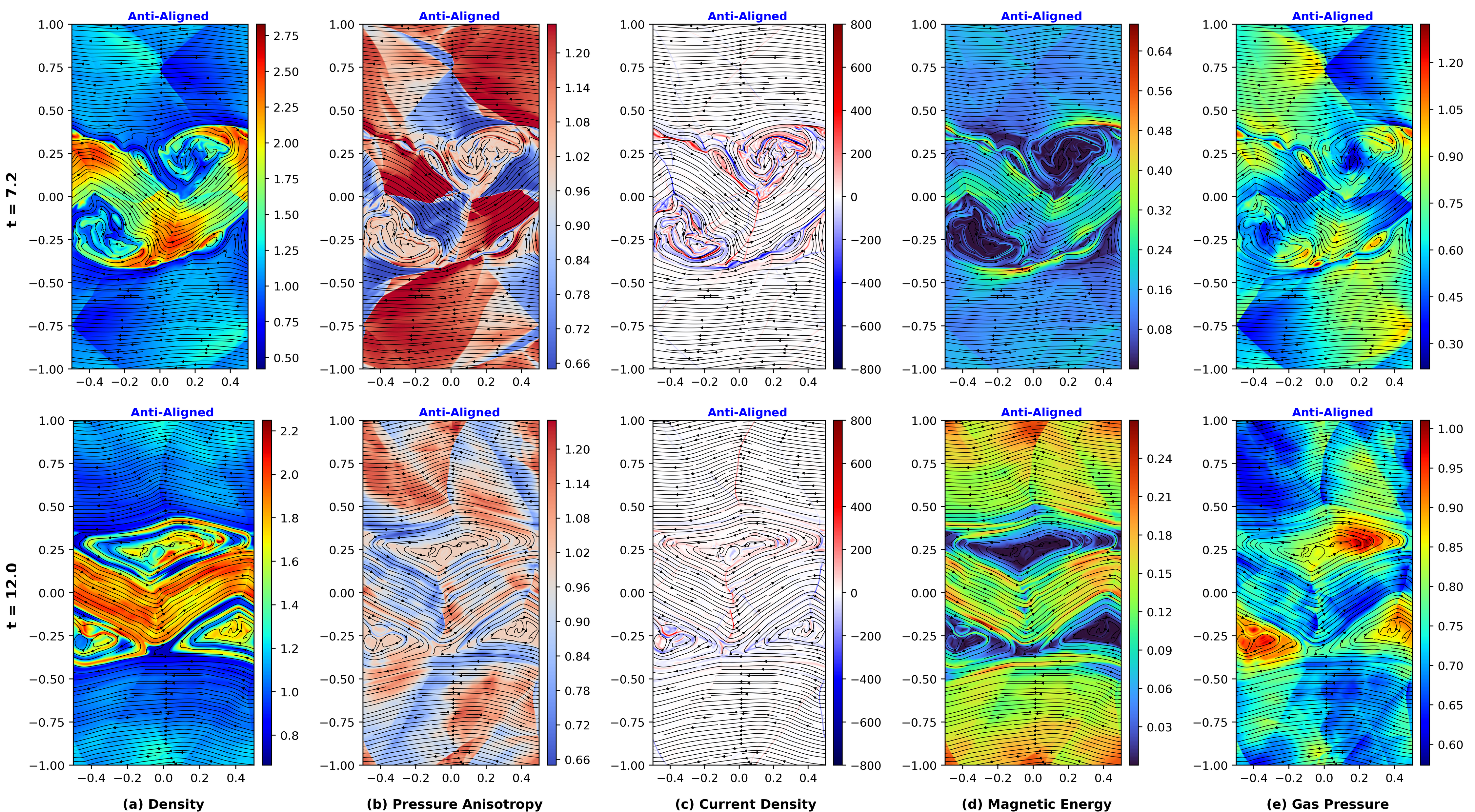}
		\caption{Density ($\rho$), pressure anisotropy ($A = p_\perp / p_\parallel$), current density ($J_z$), magnetic energy ($B_x^2 + B_y^2$), and total gas pressure $\bigl[(p_\parallel + 2p_\perp)/3\bigr]$ at early time ($t=7.2$, top row) and late time ($t=12.0$, bottom row) for the simulation with an anti-aligned magnetic field and CGL relaxation time $\tau = 5.0$.}
		\label{fig_2}
	\end{center}
\end{figure}

\begin{figure}%[H]
	\begin{center}
		\includegraphics[width=1.0\textwidth,clip=]{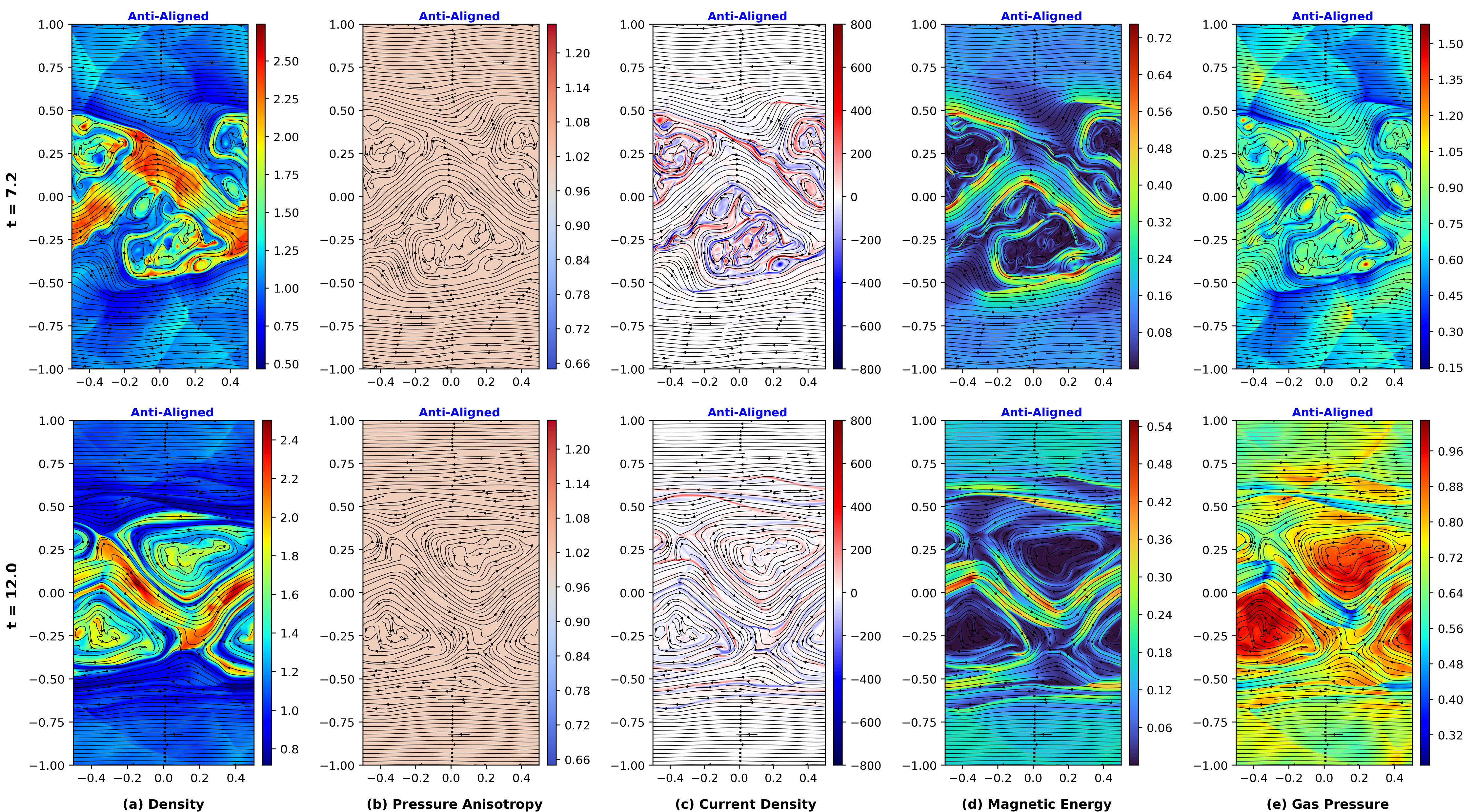}
		\caption{Same as Figure \ref{fig_2}, but for a CGL relaxation time $\tau = 10^{-5}$, which corresponds to the MHD limit.}
		\label{fig_3}
	\end{center}
\end{figure}

\begin{figure}%[H]
	\begin{center}
		\includegraphics[width=1.0\textwidth,clip=]{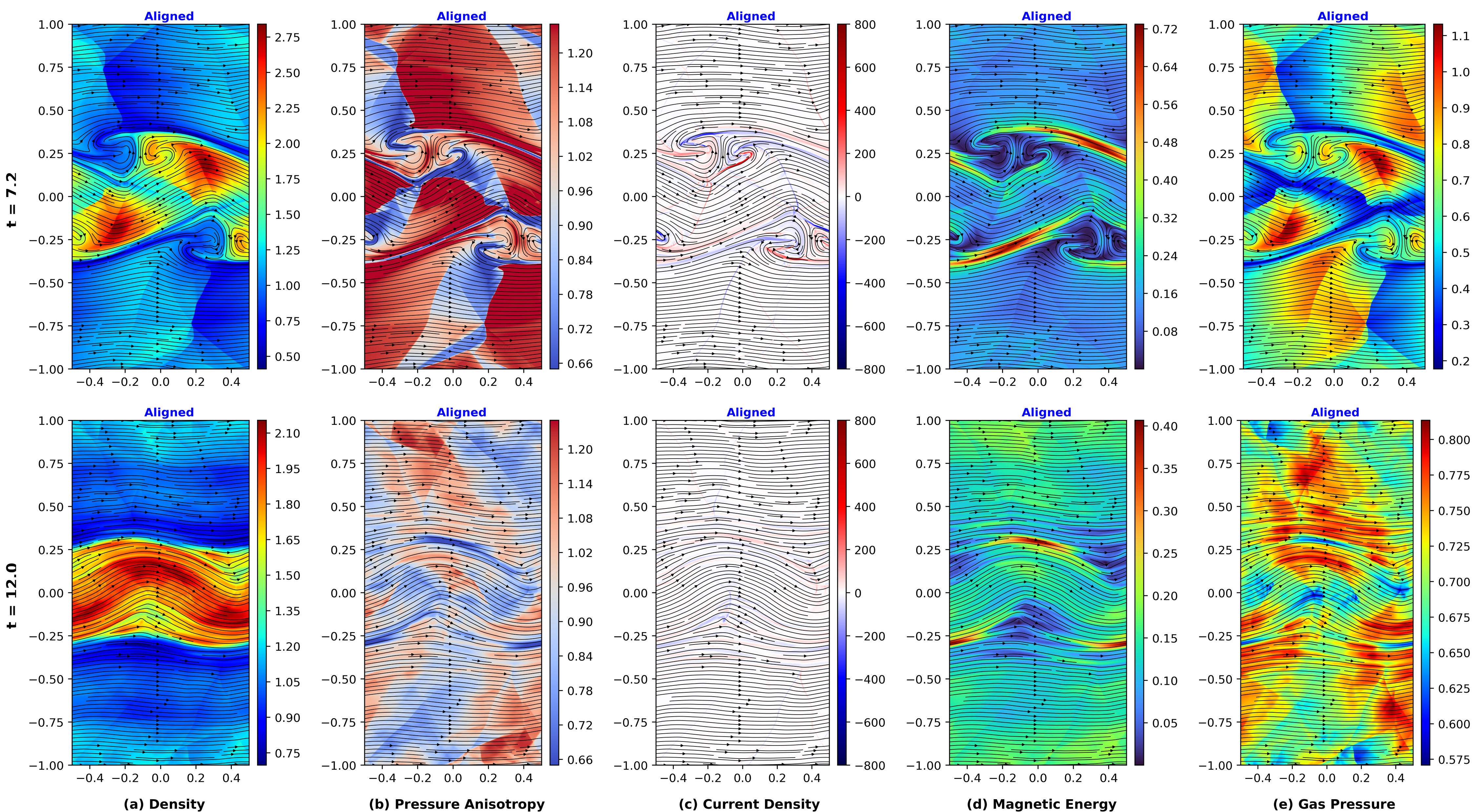}
		\caption{Same as Figure \ref{fig_2}, but for the aligned magnetic field simulation.}
		\label{fig_4}
	\end{center}
\end{figure}

\begin{figure}%[H]
	\begin{center}
		\includegraphics[width=1.0\textwidth,clip=]{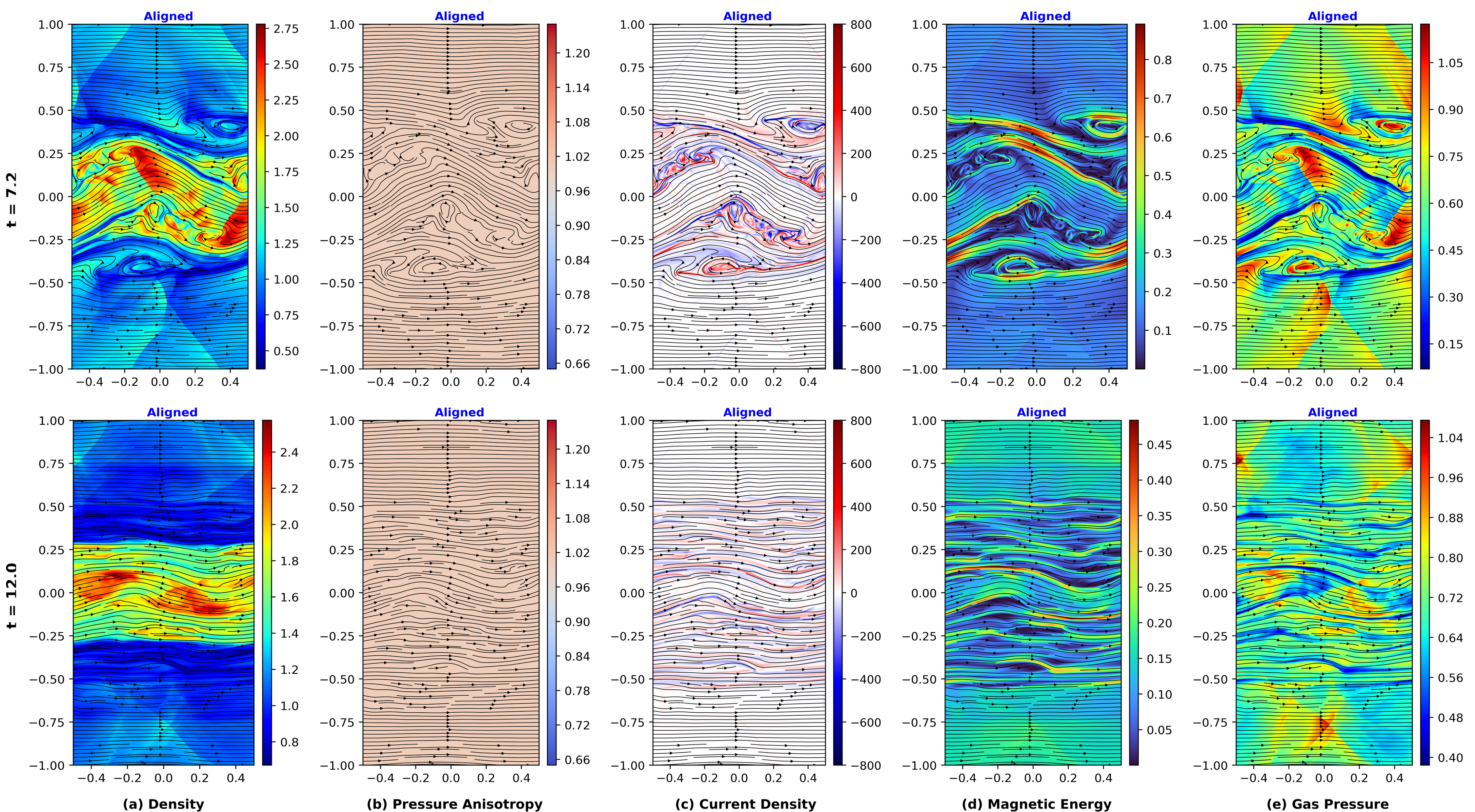}
		\caption{Same as Figure \ref{fig_4}, but for a CGL relaxation time $\tau = 10^{-5}$, which corresponds to the MHD limit.}
		\label{fig_5}
	\end{center}
\end{figure}

We begin by focusing on Fig. \ref{fig_2} which shows the development of the KH instability for the situation where the magnetic field is anti-aligned. In Fig. \ref{fig_2}, we show a simulation that is highly susceptible to the development of pressure anisotropies because the relaxation time $\tau$ has been set to $5.0$. The upper panel in Fig. \ref{fig_2} corresponds to a time of $7.2$ which we identify as a time when the linear growth is about to end. The lower panel in Fig. \ref{fig_2} shows the results at a time of $12.0$ by which time the full non-linear growth is apparent. The choice of a time of $7.2$ as a time when the linear growth is about to end will be justified in the next Section where we make quantitative analysis of the simulated data. From left to right, Fig. \ref{fig_2} shows the density, the anisotropy ratio ``$A$'', the current density, the magnetic energy, and the mean pressure $(p_\parallel + 2p_\perp)/3$. To facilitate intercomparison across plots that show the simulated data, we have shown the anisotropy ratio on the same scale in all plots. Likewise, the current density is a good indicator of the amount of dissipation due to reconnection events and, as a result, to facilitate intercomparison across plots, we will show it on the same scale. In this Section we will be very interested in making visual intercomparisons between situations where anisotropy can develop, and the MHD limit where anisotropy cannot develop. For that reason, Fig. \ref{fig_3} shows data from a simulation that is almost identical to the one shown in Fig. \ref{fig_2} with the exception that the relaxation time for the simulation shown in Fig. \ref{fig_3} is $10^{-5}$. Similarly, we will be interested in comparing KH instability evolution where the magnetic field is anti-aligned to situations where the magnetic field is aligned. For that reason, Fig. \ref{fig_4} shows results from a simulation with relaxation time of $5.0$ but this time the magnetic fields shown in Fig. \ref{fig_1}a are aligned. Fig. \ref{fig_5} shows data from a simulation that is almost identical to the one shown in Fig. \ref{fig_4} with the exception that the relaxation time for the simulation shown in Fig. \ref{fig_5} is $10^{-5}$. 

% =====================================
% Figure 2: Tau = 5 (Anti-Aligned Case)
% =====================================

In Fig. \ref{fig_2}a we see that the density variable in the central flow channel, which has denser fluid, has undergone increasing amounts of deformation as time progresses. This behavior is also matched by the density variable in Fig. \ref{fig_3}a. Owing to the much larger relaxation time in Fig. \ref{fig_2}b, we can see that the anisotropy also develops to substantial levels as time progresses. At early times, we see the development of very prominent anisotropy ratios that are much larger than unity at the boundaries of the central channel, indicating that the KH instability applies compressive fluctuations to the field lines, which is known to lead to a strong mirror instability. At later times, we see that shocks that develop in the region that is exterior to the channel have produced compressions as well as field line stretching leading to regions with $A > 1$, i.e. mirror instability, as well as regions with $A < 1$, i.e. firehose instability. Now please compare the two panels in Fig. \ref{fig_2}b to the two panels in Fig. \ref{fig_3}b. In Fig. \ref{fig_3}b we see practically no anisotropy in the pressure at early or late times. This is entirely consistent with the very small relaxation time for the simulation shown in Fig. \ref{fig_3}b. While we have not shown it here, if we were to show the other simulations from Table \ref{table_1_Simulation_Runs} with relaxation times of $1.0$ and $0.1$ they would show the same trends.

% =========================================
% Figure 3: Tau = 10e-5 (Anti-Aligned Case)
% =========================================

The previous paragraph has cross-compared density and anisotropy ratios between Figs. \ref{fig_2} and \ref{fig_3}. We now cross-compare variables in Figures \ref{fig_2} and \ref{fig_3} that seem to display a stronger dependence on the magnetic field. Please consider the upper and lower panels in Fig. \ref{fig_2}c for the current density. We see that the current density peaks early and then settles down to lower values. This suggests that in the anti-aligned case the substantial destruction of magnetic energy takes place early-on. Now please cross-compare the upper and lower panels in Fig. \ref{fig_2}c to the upper and lower panels in Fig. \ref{fig_3}c. Since the currents have been shown on the same scale in Figs. \ref{fig_2}, \ref{fig_3}, \ref{fig_4} and \ref{fig_5}, we see quite clearly that the regions with strong currents fill a larger volume in Fig. \ref{fig_3}c than in Fig. \ref{fig_2}c. This indicates a greater propensity for reconnection in the MHD limit than in the CGL limit. Our hunch that this is so is also borne out when we cross compare the magnetic energies in Fig. \ref{fig_2}d with the magnetic energies in Fig. \ref{fig_3}d. The dark magnetic islands in Figs. \ref{fig_2}d and \ref{fig_3}d show the regions where substantial magnetic field destruction has occurred. The distance between magnetic islands is also smaller in Fig. \ref{fig_3}d than in Fig. \ref{fig_2}d, indicating that more magnetic energy has been destroyed in the simulation reported in Fig. \ref{fig_3}d. We see that substantially more magnetic energy was destroyed in the MHD limit than in the CGL limit. Figs. \ref{fig_2}e and \ref{fig_3}e show that the regions where magnetic energy, and therefore magnetic pressure, is diminished are also locations where the gas pressure has compensated in order to maintain overall pressure balance. Therefore, we see larger patches of red in Fig. \ref{fig_3}e than in Fig. \ref{fig_2}e. This observation will also be borne out in the analysis in the next Section. While we have not shown it here, if we were to show the other simulations with relaxation times of $1.0$ and $0.1$ they would show the same trends. It is fair to ask why the MHD limit shows a greater propensity for reconnection than the CGL limit? We offer one intuitive explanation here. The CGL simulations can access two pressure channels, i.e. the parallel and perpendicular pressures. As a result, turbulent energy that develops in the non-linear phase of growth can be accommodated in (and shuttle between the) two sets of modes in the CGL limit. This means that less energy goes into the Alfv\'{e}nic modes that might contribute to reconnection. In the MHD limit, i.e. in Fig. \ref{fig_3}, the simulation cannot access these additional modes, and as a result, reconnection is enhanced.

% =====================================
% Figure 4: Tau = 5 (Aligned Case)
% =====================================

In the previous two paragraphs we intercompared Figs. \ref{fig_2} and \ref{fig_3} which showed the trends in the KH instability with anti-aligned magnetic fields in the CGL and MHD limits respectively. We are very interested in asking whether the some of the same trends carry over to the KH instability with aligned magnetic fields in the CGL and MHD limits? Consider Fig. \ref{fig_4} which shows the KH instability simulated with aligned magnetic field when the relaxation time is set to $5.0$ units; i.e. the CGL limit. Fig. \ref{fig_5} shows analogous information for the KH instability simulated with aligned magnetic field when the relaxation time is set to $10^{-5}$ units; i.e. the MHD limit. Comparing the densities in Figs. \ref{fig_4}a and \ref{fig_5}a, we see that the flow channel in Fig. \ref{fig_4}a has a much more coherent density structure compared to Fig. \ref{fig_5}a where the density at final times is much more patchy. This shows us at the very onset that much more turbulent action has taken place in Fig. \ref{fig_5}a than in Fig. \ref{fig_4}a. Fig. \ref{fig_4}b shows that we have substantial amounts of anisotropy forming in the early stages of the KH instability which subsides somewhat as time progresses. Fig. \ref{fig_5}b shows no anisotropy formation, as expected. The trends seen at early and late times in Fig. \ref{fig_4}b also mirror those seen in Fig. \ref{fig_2}b. Comparison of Fig. \ref{fig_4}c and Fig. \ref{fig_5}c is also very interesting because it shows that the MHD simulation has a greater propensity towards larger and more pervasive current densities than the CGL simulation. This is also the same trend that we saw developing between Figs. \ref{fig_2}c and \ref{fig_3}c. Comparison of Figs. \ref{fig_4}d and \ref{fig_5}d shows that the magnetic energy in Fig. \ref{fig_4}d is larger and more evenly spread out than in Fig. \ref{fig_5}d. Fig. \ref{fig_5}d also shows the formation of magnetic vortices in the flow; a trend that is not apparent in Fig. \ref{fig_4}d. We see that the trends in magnetic energy anti-correlate with the trends in the current density. It is also significant that the trend seen between Figs. \ref{fig_2}d and \ref{fig_3}d are also mirrored between Figs. \ref{fig_4}d and \ref{fig_5}d. Of course, Figs. \ref{fig_4}d and \ref{fig_5}d do not show the formation of large magnetic islands that we saw in Figs. \ref{fig_2}d and \ref{fig_3}d; but this can easily be attributed to the aligned magnetic configuration in Figs. \ref{fig_4} and \ref{fig_5}. The magnetic pressure variations in Fig. \ref{fig_5}e also have a larger range than the magnetic pressure variations in Fig. \ref{fig_4}e.

% =========================================
% Figure 5: Tau = 10e-5 (Aligned Case)
% =========================================

%%%%%%%% Section:5 Diagnostics (growth rate, intermittency & Brazil plot) %%%%%%%%

\section{Diagnostics: Measurement of growth rate, intermittency \& Brazil plot} \label{sec:diagnostics}

%%%%%%%% Sub-Section:5.1 Comparison of Numerical \& Analytical Growth Rates %%%%%%%%
\subsection{Comparison of Numerical \& Analytical Growth Rates}\label{subsec:growth rate}
\begin{figure}%[H]
	\begin{center}
		\includegraphics[width=0.85\textwidth,clip=]{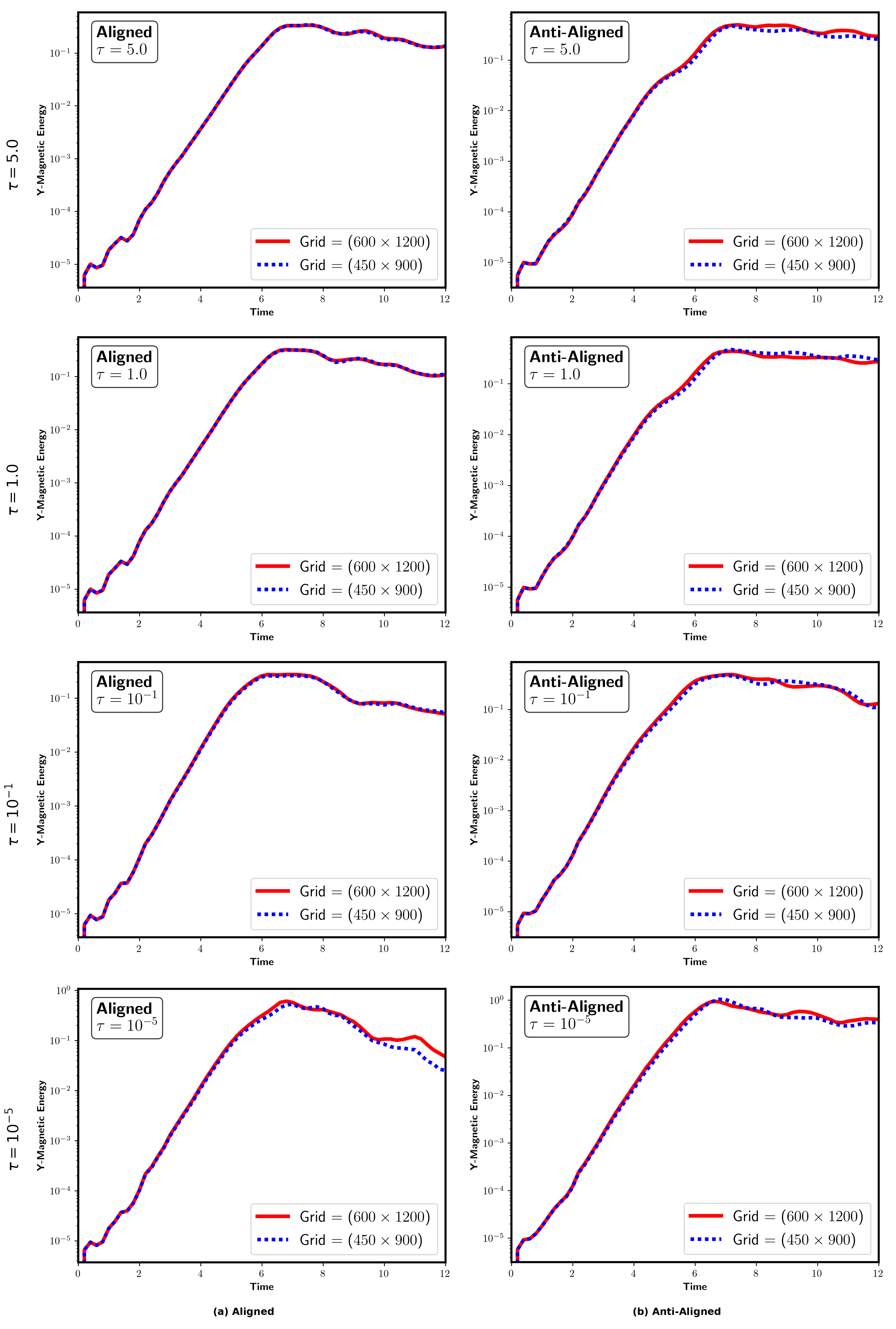}
		\caption{Evolution of the spatially-averaged transverse magnetic energy as function of time for different CGL relaxation times ($\tau = 5.0$, $1.0$, $0.1$, \& $10^{-5}$) in both aligned and anti-aligned magnetic field cases. The numerical growth rates, which are compared with the linearised ones (refer to Table \ref{table_2_aligned} \& \ref{table_3_anti_aligned}), are extracted from these plots. The solid red line corresponds to a grid resolution of $600 \times 1200$, and the blue dashed line corresponds to a grid resolution of $450 \times 900$. We observe very good convergence between the two resolutions, and the indistinguishable growth of magnetic energy confirms the numerical convergence of the results.}
		\label{fig_6}
	\end{center}
\end{figure}

We now proceed to compare the numerical and linearised growth rates of the KH instability for different CGL relaxation times ($\tau = 5.0$, $1.0$, $0.1$, \& $10^{-5}$), considering both aligned and anti-aligned magnetic field configurations. To this end, we compute the transverse magnetic energy (i.e., the spatially-averaged value of the $y$-component of magnetic energy) as a function of time (see Fig. \ref{fig_6}). In the log-linear representation of the transverse magnetic energy, as shown in Fig. \ref{fig_6}, the exponential growth [$\exp (2\Gamma t)$] phase manifests itself as a straight line. By performing a linear fit to this early portion of the curve, we estimate the numerical growth rate ($\Gamma$) of the KH instability for each case. A detailed comparison between the numerical and linearised growth rates is presented in Table \ref{table_2_aligned} for the aligned magnetic field configuration and in Table \ref{table_3_anti_aligned} for the anti-aligned configuration. From Tables \ref{table_2_aligned} and \ref{table_3_anti_aligned}, we see that the numerical growth rates, extracted from the exponential growth of the transverse magnetic energy, closely match the corresponding linearised predictions for both aligned and anti-aligned magnetic field configurations. This demonstrates the consistency between our simulations and linearised expectations. It is important to note that, for both aligned and anti-aligned magnetic field configurations, the growth rates of the KH instability are suppressed in the CGL limit ($\tau = 5.0$) compared to the MHD limit ($\tau = 10^{-5}$), as evident from both numerical and linearised predictions. The rather nice match-up between the numerical and linearised growth rates also provides a validation that the code is producing the correct results. The growth rates for the aligned simulations in Table \ref{table_2_aligned} match slightly better than the growth rates for the anti-aligned simulations in Table \ref{table_3_anti_aligned} because the anti-aligned cases are more prone to the effects of numerical resistivity than the aligned cases. The simulation is run until $t = 12.0$, which is equivalent to approximately $24$ eddy turnover times. As illustrated in Figure \ref{fig_6}, the magnetic energy experiences a exponential growth during the linear phase, but it attains a plateau at approximately $t = 7.0$. A comparable pattern is observed in the evolution of kinetic energy. This implies that the system has achieved a statistically stable state by the time $t \sim 7$.

% =========================================
% Figure 6: Y-Magnetic Energy
% =========================================

% =======================================================
% Table 2: Growth rates numerical vs analytical (aligned)
% =======================================================
\begin{table}%[H]
\begin{center}
\caption{A comparison between linearised and numerical growth rates ($\Gamma$) at different CGL relaxation times ($\tau = 5.0$, $1.0$, $0.1$, and $10^{-5}$) for the KH instability with an aligned magnetic field.}
\label{table_2_aligned}
\begin{tabular}{cccccccccc}
\tableline
             & $\tau = 5.0$   &   $\tau = 1.0$     & $\tau = 0.1$  & $\tau = 10^{-5}$ \\
\tableline
Analytical   & $0.978$        &    $1.013$         &     $1.160$   &       $1.167$   \\
Numerical    & $0.977$        &    $1.013$         &     $1.161$   &       $1.166$     \\
\tableline
\end{tabular}
\end{center}
\end{table}

% ============================================================
% Table 3: Growth rates numerical vs analytical (anti-aligned)
% ============================================================
\begin{table}%[H]
\begin{center}
\caption{Same as Table \ref{table_2_aligned}, but for the anti-aligned magnetic field.}
\label{table_3_anti_aligned}
\begin{tabular}{cccccccccc}
\tableline
             & $\tau = 5.0$    &   $\tau = 1.0$  & $\tau = 0.1$  & $\tau = 10^{-5}$ \\
\tableline
Analytical   & $1.033$         &    $1.063$      &     $1.223$   &       $1.105$   \\
Numerical    & $1.098$         &    $1.094$      &     $1.194$   &       $1.171$     \\
\tableline
\end{tabular}
\end{center}
\end{table}

%%%%%%%% Sub-Section:5.2 Anisotropy Evolution in CGL %%%%%%%%
\subsection{Anisotropy Evolution in CGL}\label{subsec:anisotropy}
\begin{figure}%[H]
	\begin{center}
		\includegraphics[width=1.0\textwidth,clip=]{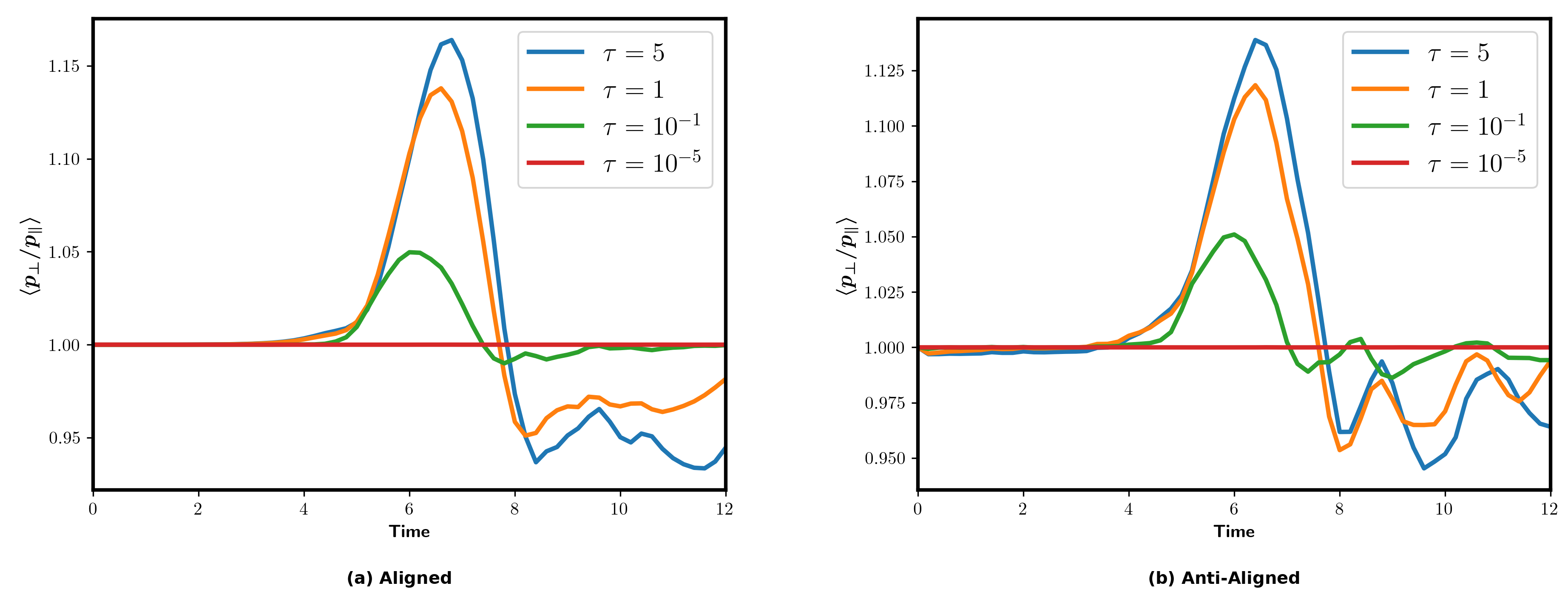}
		\caption{Evolution of the spatially-averaged pressure anisotropy ($A = \langle p_\perp/p_\parallel\rangle$) as a function of time for both aligned and anti-aligned magnetic field cases at different CGL relaxation times ($\tau = 5.0$, $1.0$, $0.1$, and $10^{-5}$).}
		\label{fig_7}
	\end{center}
\end{figure}

Following the comparison of numerical and linearised growth rates, we now turn our attention to the evolution of pressure anisotropy, which plays a crucial role in regulating the nonlinear dynamics of the KH instability in the CGL framework.
Here, we examine the evolution of pressure anisotropy during the development of the KH instability in the CGL framework. Figure \ref{fig_7} shows the temporal evolution of the spatially averaged pressure anisotropy, expressed as $A = \langle p_\perp/p_\parallel\rangle$, for both aligned (refer to Fig. \ref{fig_7}a) and anti-aligned (refer to Fig. \ref{fig_7}a) magnetic field configurations, and for different CGL relaxation times ($\tau = 5.0$, $1.0$, $0.1$, and $10^{-5}$). In all cases, the ratio initially rises above unity during the early (linear) phase of the instability and subsequently decreases below unity as the system transitions into the nonlinear stage. This behavior suggests the distinct roles of mirror and firehose instabilities during different phases of the KH evolution. In the linear regime, the formation and roll-up of KH fluctuations compress the magnetic field lines. According to the double-adiabatic (CGL) invariants, such compression enhances the perpendicular pressure ($p_\perp$) while reducing the parallel pressure ($p_\parallel$) leading to $p_\perp > p_\parallel$. This pressure anisotropy drives the plasma toward the mirror instability threshold. As the system evolves further into the nonlinear regime, the KH flow causes the field lines to make sinusoidal oscillations. This favors the growth of $p_\parallel$ relative to $p_\perp$, eventually driving the system towards the firehose instability threshold. Consequently, mirror-type fluctuations should dominate during the linear growth phase, whereas firehose-like signatures would emerge predominantly in the nonlinear phase of the instability.

% =============================================
% Figure 7: Avg. of Pressure anisotropy vs time
% =============================================

\begin{figure}%[H]
	\begin{center}
		\includegraphics[width=1.0\textwidth,clip=]{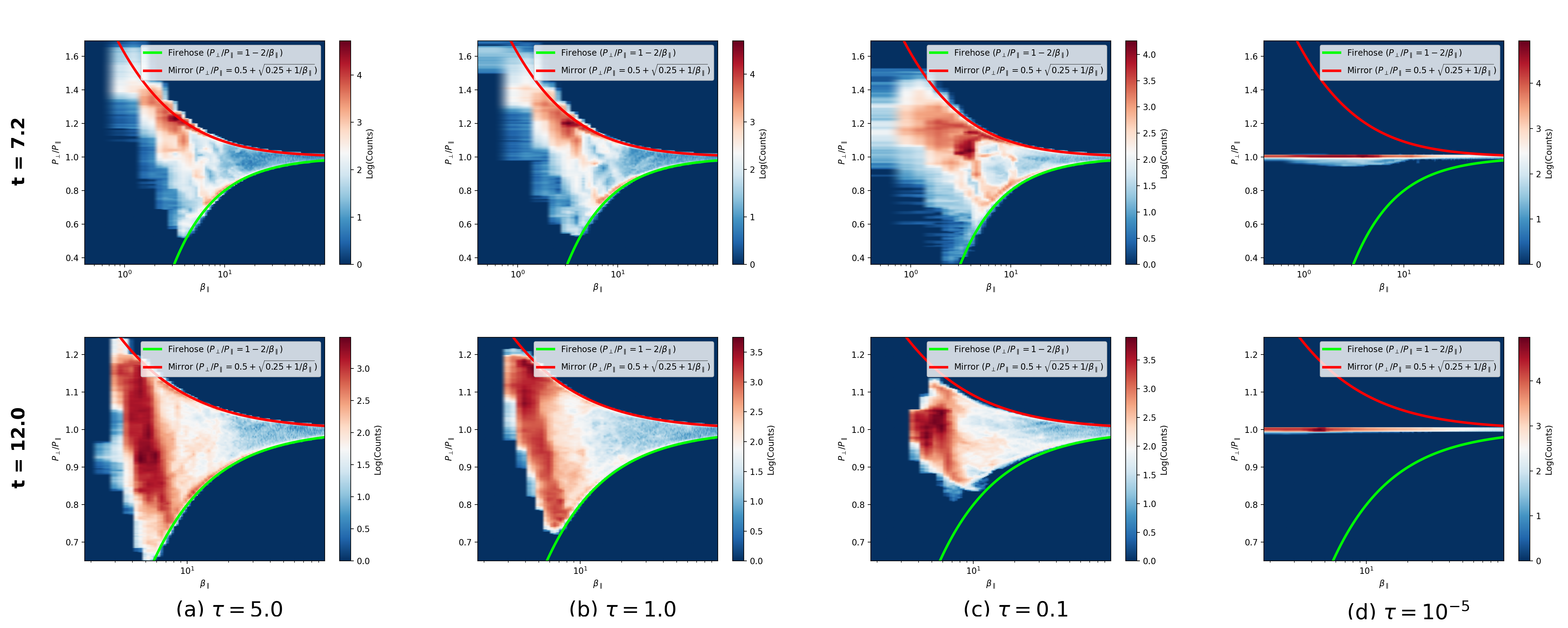}
		\caption{Binned distributions of pressure anisotropy values ($A = p_\perp/p_\parallel$) as a function of the parallel plasma beta ($\beta_\parallel$) at early time ($t = 7.2$, top row) and late time ($t = 12.0$, bottom row) for the KH instability with an anti-aligned magnetic field, at different CGL relaxation times ($\tau = 5.0$, $1.0$, $0.1$, and $10^{-5}$). The green lines indicate the firehose instability threshold (refer to Equation \ref{eq:firehose}), while the red lines indicate the mirror instability (refer to Equation \ref{eq:mirror}) threshold.}
		\label{fig_8}
	\end{center}
\end{figure}

The pressure anisotropy (aka Brazil) plots \citep{Bale_2009} in Figure \ref{fig_8} depict the binned distribution of pressure anisotropy values ($A = p_\perp/p_\parallel$) as a function of the parallel plasma beta ($\beta_\parallel$) for different CGL relaxation times ($\tau = 5.0, 1.0, 0.1, 10^{-5}$), shown at both early and late times. The top panel of Fig. \ref{fig_8} corresponds to an early time, while the bottom panel represents a later time. In the early, linear phase of the KH instability, the compression of magnetic field lines in the shear layer preferentially increases the perpendicular pressure, leading to $p_\perp > p_\parallel$. This is evident in the Brazil plots, where the majority of counts cluster near the mirror instability threshold, given by 

\begin{align}
	A \equiv \frac{p_\perp}{p_\parallel} &= 0.5 + \sqrt{0.25+\frac{1}{\beta_\parallel}}
	\label{eq:mirror}
\end{align}
highlighting that mirror modes dominate the linear dynamics (refer to the upper row of Fig. \ref{fig_8} ).

% =============================================
% Figure 8: Brazil Plot (Anti-Aligned)
% =============================================

As the system evolves into the nonlinear regime, the magnetic field lines begin to relax and straighten, causing some regions to become susceptible to the firehose instability, defined by

\begin{align}
	A \equiv \frac{p_\perp}{p_\parallel} &= 1.0 - \frac{2.0}{\beta_\parallel}.
	\label{eq:firehose}
\end{align}
This redistribution of pressure anisotropy leads to a broader filling of the pressure space, with significant counts now spanning the region between the mirror and firehose thresholds (refer to the bottom row of Fig. \ref{fig_8} ). Such a distribution demonstrates that, in the nonlinear stage, both mirror and firehose instabilities should contribute to the plasma dynamics, based on the temporal evolution of $\langle p_\perp/p_\parallel\rangle$ observed in the simulations (refer to Fig. \ref{fig_7}). Here, we show only the anti-aligned cases; the aligned configurations follow the same trend and are therefore omitted for the sake of brevity.

As seen in Fig. \ref{fig_8}, in the CGL limit (i.e., for higher relaxation times $\tau = 5.0$), the histogram of $p_\perp / p_\parallel$ is broadly populated, reflecting the system's ability to sustain significant pressure anisotropy. This allows the plasma to explore a wide range of values between the mirror and firehose instability thresholds. In contrast, in the MHD limit (i.e., for very low relaxation times $\tau = 10^{-5}$), the pressure rapidly relaxes toward isotropy, and the histogram counts are sharply concentrated around $A = p_\perp / p_\parallel \approx 1$. This behavior is consistent with the expected physics: in the CGL limit, anisotropy-driven instabilities can develop, whereas in the MHD limit, fast relaxation keeps the plasma nearly isotropic.

%%%%%%%% Sub-Section:5.3 Evolution of Magnetic Energy \& Current %%%%%%%%
\subsection{Evolution of Magnetic Energy \& Current} \label{subsec: mag_energy_current}
\begin{figure}%[H]
	\begin{center}
		\includegraphics[width=1.0\textwidth,clip=]{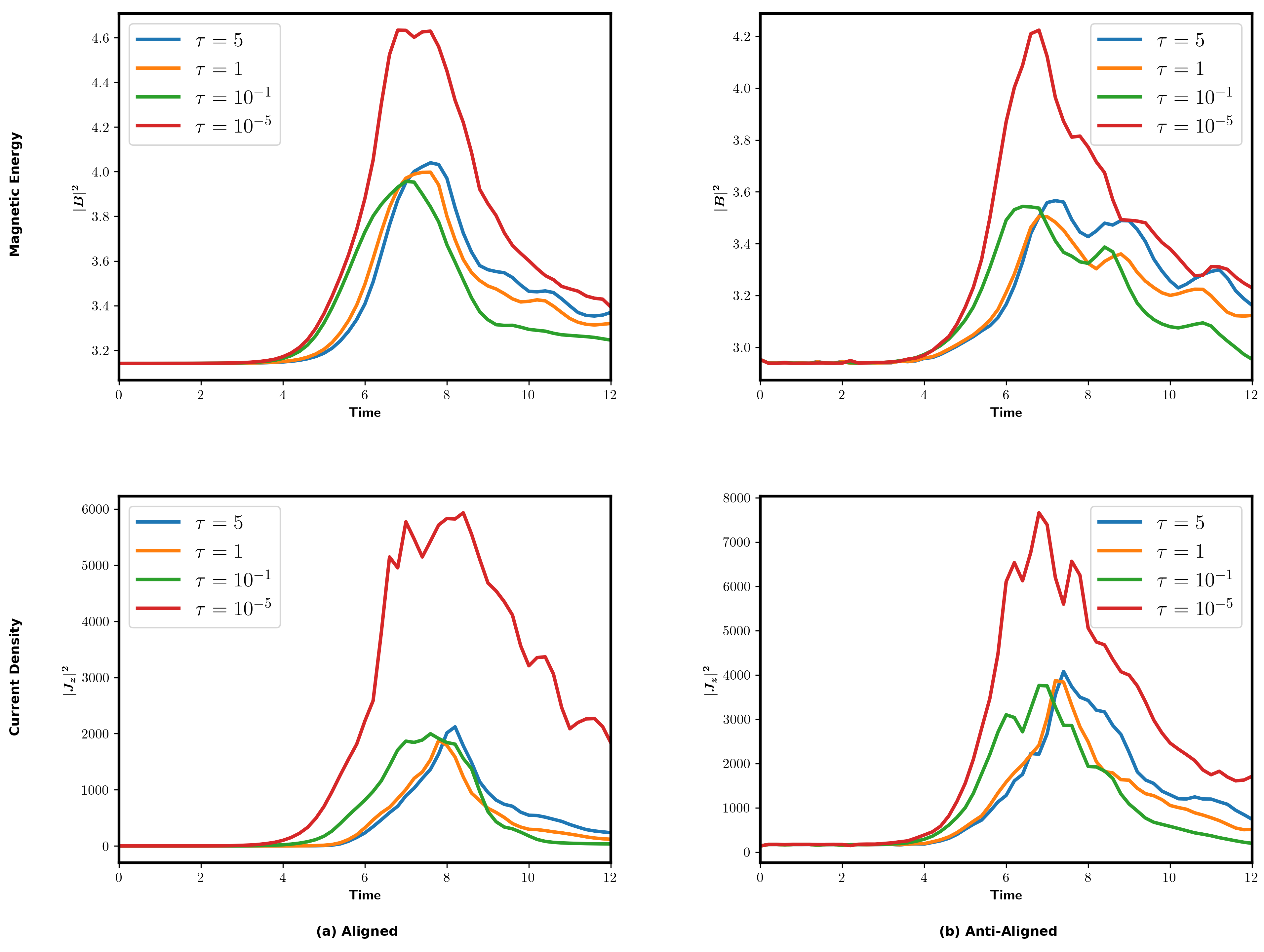}
		\caption{Evolution of the spatially averaged magnetic energy ($|B|^2$) [top row] and the current density ($|J_z|^2$) [bottom row] as a function of time for both aligned and anti-aligned magnetic field cases, at different CGL relaxation times ($\tau = 5.0$, $1.0$, $0.1$, and $10^{-5}$). Note that magnetic reconnection occurs more rapidly in the MHD limit (i.e., $\tau = 10^{-5}$ case).}
		\label{fig_9}
	\end{center}
\end{figure}

In this Sub-section, we analyze the temporal evolution of the spatially averaged magnetic energy [top row of Fig. \ref{fig_9}] and current density [bottom row of Fig. \ref{fig_9}] for both aligned and anti-aligned magnetic field configurations across different CGL relaxation times ($\tau = 5.0$, $1.0$, $0.1$, and $10^{-5}$). These diagnostics provide insight into the development of the KH instability and the associated magnetic reconnection processes in our simulations. In this study, we intentionally focus on the evolution governed solely by the native set of equations, without incorporating any anomalous resistivity. Because CGL simulations represent a relatively unexplored regime, our aim here is to first understand the fundamental dynamics that arise in the absence of such additional dissipation mechanisms.

% =============================================
% Figure 9: Magnetic Energy and Current vs time
% =============================================

The evolution of the spatially-averaged magnetic energy in our simulations exhibits a clear dependence on the CGL relaxation time. As shown in Fig. \ref{fig_9} (top row) , for both aligned and anti-aligned magnetic field cases, the magnetic energy reaches a higher peak and declines more rapidly in the MHD limit (i.e., very low relaxation time, $\tau = 10^{-5}$) compared to the CGL limit ($\tau = 5.0$). Physically, this observation reflects the enhanced efficiency of magnetic reconnection in the MHD regime relative to the CGL regime. In the CGL regime, part of the energy can be partitioned to generate pressure anisotropies, whereas in the MHD regime the same energy is instead utilized for bending magnetic field lines. The stronger bending of magnetic field lines in the MHD regime channels more energy into  Alfv\'{e}nic modes, which in turn enhances magnetic reconnection. Consequently, the MHD cases exhibit more efficient magnetic reconnection compared to the CGL cases.

In the CGL framework, the plasma possesses two distinct pressure channels, parallel ($p_\parallel$) and perpendicular ($p_\perp$) to the magnetic field. Consequently, turbulent energy that develops during the non-linear phase of the KH instability can be accommodated within these two sets of modes and can shuttle between them. This redistribution of energy implies that a smaller fraction of the available energy is channeled into Alfv\'{e}nic modes that are directly responsible for stretching, bending, and reconnecting magnetic field lines. As a result, magnetic reconnection is less efficient in the CGL limit, leading to a slower growth and decay of magnetic energy (refer to the top row of Fig. \ref{fig_9}). A similar conclusion has been reached by \citet{Ferreira-Santos_2025} in their study of reconnection in CGL plasmas; though those authors were restricted to having a non-zero guide field because of a limitation of their numerical methods. This present study shows that even without a guide field, and even when other effects such as KH instabilities are also operative, the reduced reconnection in CGL plasmas (vis a vis MHD) holds.

By contrast, in the MHD limit, where the pressure is isotropic and no separate parallel or perpendicular pressure channels exist, the turbulent energy is more directly available to drive Alfv\'{e}nic motions. Without the additional degrees of freedom for energy redistribution, reconnection proceeds more efficiently, producing a higher peak in magnetic energy and a faster decay of magnetic energy (refer to the top row of Fig. \ref{fig_9}). %These results appear to contradict some steady anisotropic reconnection simulations \citep[e.g.][]{Hirabayashi_Hoshino_2013}, that suggest that the rate of magnetic reconnection is slightly increased in the CGL approximation compared to MHD, owing to a decrease in perpendicular pressure leading to a faster inflow into the current sheet. However, reconnection in the presence of velocity shear is radically different from the steady-state case, and we instead observe the dominance of the perpendicular pressure in the earlier phases of the instability development, that would prevent the inflow into the current sheet and hinder magnetic reconnection.%

The evolution of the spatially-averaged squared current density, $J_z^2$, shown in the lower panel of Fig. \ref{fig_9}, exhibits trends consistent with those observed for the magnetic energy. For both aligned and anti-aligned magnetic field configurations, $J_z^2$ shows a higher growth followed by a faster decay in the MHD limit (lower CGL relaxation time $\tau = 10^{-5}$) compared to the CGL limit (higher CGL relaxation time $\tau = 5.0$). This behavior further supports the conclusion that magnetic reconnection is more efficient in the MHD regime. In the CGL limit, a portion of the energy is accommodated within the parallel and perpendicular pressure channels, leading to a slower growth and more gradual decay of $J_z^2$. These observations reinforce the critical role of pressure anisotropy in modulating the nonlinear dynamics of the KH instability.

% ============================================================
% Figure 10: Y-Magnetic Energy vs time aligned vs anti-aligned
% ============================================================
\begin{figure}%[H]
	\begin{center}
		\includegraphics[width=1.0\textwidth,clip=]{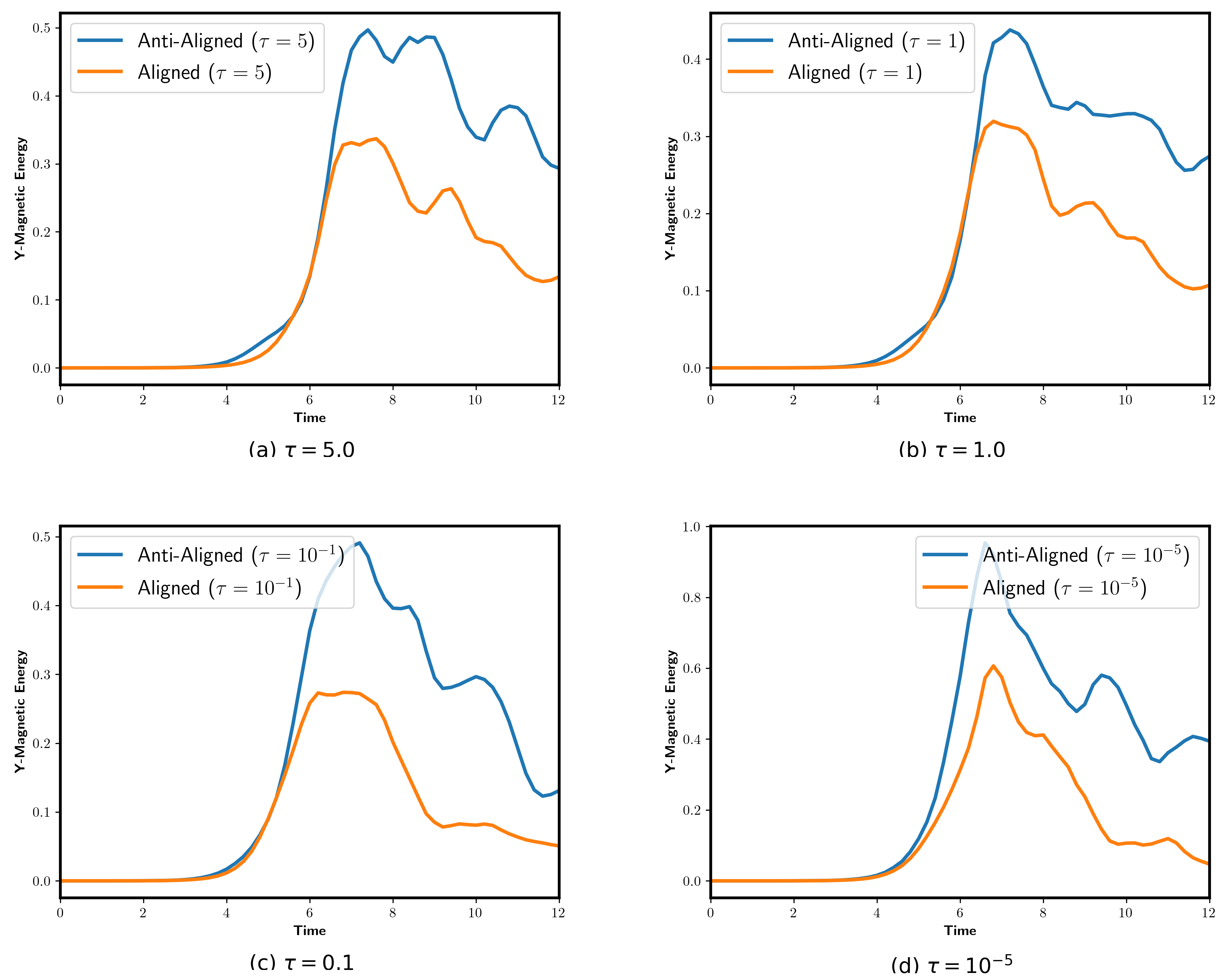}
		\caption{Evolution of the spatially averaged transverse magnetic energy as a function of time at different CGL relaxation times ($\tau = 5.0$, $1.0$, $0.1$, and $10^{-5}$), for both aligned and anti-aligned magnetic field cases.}
		\label{fig_10}
	\end{center}
\end{figure}

This behavior is consistent with our earlier observations from the visual data presented in Figs. \ref{fig_2}c, \ref{fig_2}d, \ref{fig_3}c, \ref{fig_3}d, \ref{fig_4}c, \ref{fig_4}d, \ref{fig_5}c, and \ref{fig_5}d. In the CGL limit, a portion of the energy can be temporarily stored in anisotropy-driven pressure modes, which slows down the magnetic reconnection process. In contrast, in the MHD limit, the plasma remains nearly isotropic, and the absence of these additional anisotropic channels allows energy to be restricted more efficiently into the development of magnetic islands and current sheets, resulting in a higher growth rate followed by a faster decay of magnetic energy and current density.

To facilitate a direct comparison, it is also instructive to present the results from the aligned and anti-aligned magnetic field configurations together for different CGL relaxation times ($\tau = 5.0$, $1.0$, $0.1$, and $10^{-5}$), as shown in Fig. \ref{fig_10}. This side-by-side representation allows us to clearly assess the similarities and differences in their dynamical evolution.
    
%%%%%%%% Sub-Section:5.4 Intermittency %%%%%%%%
\subsection{Intermittency}\label{subsec:intermittency}
Having discussed the overall dynamics, we now turn our attention to the intermittent nature of the fluctuations. Although intermittency is most meaningfully quantified in fully three-dimensional simulations, we present here some preliminary diagnostics in our 2D simulation. These results provide only a qualitative indication of intermittent behavior and suggest that a more comprehensive study in 3D might reveal more evidence for intermittency in the system.

\begin{figure}%[H]
	\begin{center}
		\includegraphics[width=1.0\textwidth,clip=]{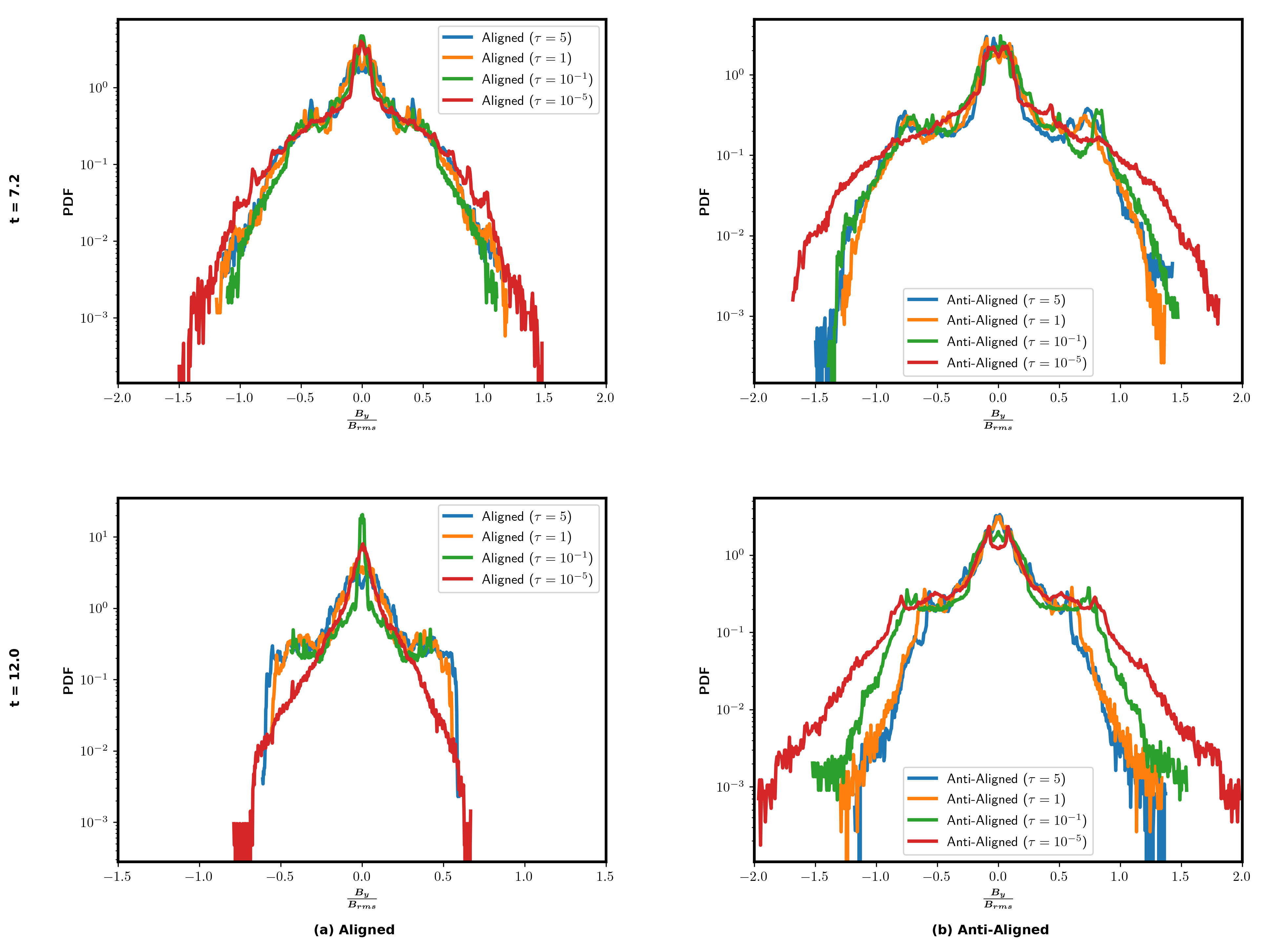}
		\caption{PDFs of the transverse magnetic field ($\frac{B_y}{B_{\rm rms}}$) at early time [top row] and late time [bottom row], for different CGL relaxation times ($\tau = 5.0$, $1.0$, $0.1$, and $10^{-5}$) in both aligned and anti-aligned magnetic field cases. The top row corresponds to time $t = 7.2$, while the bottom row corresponds to time $t = 12.0$. Note that the PDF is significantly wider for the anti-aligned case than for the aligned case.}
		\label{fig_11}
	\end{center}
\end{figure}

To investigate the possible intermittency in our simulation results, we compute the probability distribution function (PDF) of the transverse magnetic field, normalized by the root-mean-square value of the total magnetic field, $B_y/B_{rms}$, at early [top row of Fig. \ref{fig_11}] and late times [bottom row of Fig. \ref{fig_11}] for different CGL relaxation times ($\tau = 5.0, 1.0, 0.1, 10^{-5}$) in both aligned and anti-aligned magnetic field configurations. A close inspection of Fig. \ref{fig_11} reveals that the PDFs for anti-aligned cases are noticeably broader than those for aligned cases. This behavior is physically consistent with our previous observations of magnetic energy (refer to Figs. \ref{fig_2}d, \ref{fig_4}d, \ref{fig_3}d, and \ref{fig_5}d) \& current density (refer to Figs. \ref{fig_2}c, \ref{fig_4}c, \ref{fig_3}c, and \ref{fig_5}c) evolution: the anti-aligned configuration promotes stronger reconnection, leading to greater destruction of magnetic energy and the formation of magnetic islands (plasmoids). The presence of these islands introduces larger spatial variations in the transverse magnetic field, which manifests as broader PDFs (compare \ref{fig_11}a to Fig. \ref{fig_11}b). In contrast, for aligned magnetic field cases, reconnection is weaker, fewer magnetic islands form, and the resulting fluctuations are smaller, leading to narrower PDFs (refer to Fig. \ref{fig_11}a).

% =============================================
% Figure 11: PDF of By
% =============================================

It is also interesting to note that, for the aligned cases, the PDFs at late times are significantly narrower than those at early times (refer to Fig. \ref{fig_11}a), indicating a reduction in magnetic field fluctuations as the system evolves. This behavior is consistent with our earlier observations from the density and magnetic field snapshots for CGL simulations (refer to Figs. \ref{fig_4}a and \ref{fig_4}d) as well as MHD simulations (refer to Figs. \ref{fig_5}a and \ref{fig_5}d), which show that the aligned configurations gradually stabilize and develop coherent channel-like structures. In this scenario, magnetic reconnection remains weak, and the flow progressively stabilizes, leading to reduced spatial variability in $B_y$ and, consequently, narrower PDFs.

In contrast, for the anti-aligned configurations, the late-time PDFs become noticeably broader compared to their early-time counterparts (refer to Fig. \ref{fig_11}b). This trend aligns with the progressive destruction of magnetic energy and the emergence of magnetic islands observed in the density and magnetic field plots (refer to Figs. \ref{fig_2}a, \ref{fig_2}d, \ref{fig_3}a, and \ref{fig_3}d). As the system evolves into the nonlinear regime, reconnection becomes increasingly active in the anti-aligned setup, breaking the initially smooth magnetic field into fragmented plasmoid-like structures. These magnetic islands generate localized, intermittent enhancements in $B_y$, which lead to broader PDFs at later times.

% =============================================
% Figure 12: PDF of By time seris
% =============================================
\begin{figure}%[H]
	\begin{center}
		\includegraphics[width=1.0\textwidth,clip=]{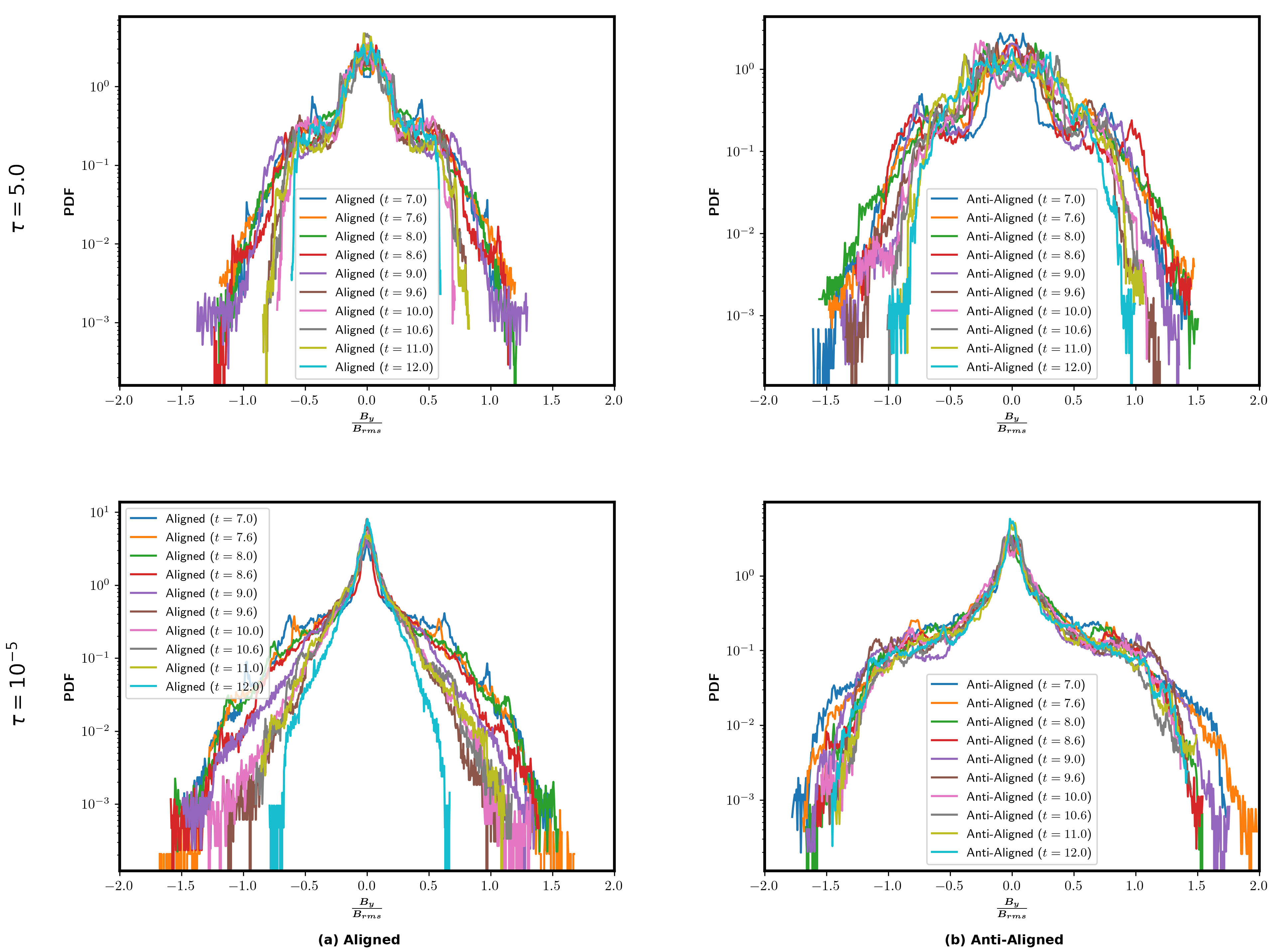}
		\caption{Time evolution of the PDFs of the transverse magnetic field ($\frac{B_y}{B_{\rm rms}}$) at CGL relaxation time $\tau = 5.0$ [top row] and $\tau = 10^{-5}$ [bottom row], for both aligned and anti-aligned magnetic field cases. The $\tau = 10^{-5}$ case (i.e., the MHD limit) exhibits a broader PDF with more extended wings than the $\tau = 5.0$ case (i.e., the CGL limit), indicating a higher possibility of intermittency.}
		\label{fig_12}
	\end{center}
\end{figure}

To further quantify the possible intermittency in our simulations, we examine the time evolution of the probability distribution function (PDF) of the transverse magnetic field, $B_y/B_{rms}$, for two representative CGL relaxation times: $\tau = 5.0$ (upper row of Fig. \ref{fig_12}) and $\tau = 10^{-5}$ (lower row of Fig. \ref{fig_12}) for both aligned and anti-aligned magnetic field configurations. The simulations are designed to have an eddy turnover time of unity, and in Fig. \ref{fig_12} we show the time variability over timescales that are smaller than that. A clear difference emerges between the two regimes: in the MHD limit ($\tau = 10^{-5}$), the PDFs exhibit pronounced extended wings, indicative of larger fluctuations and higher intermittency, whereas in the CGL limit ($\tau = 5.0$), the PDFs remain more centrally concentrated, reflecting smaller fluctuations. This behavior is physically consistent with our earlier observations of magnetic energy and current density evolution (refer to Figs. \ref{fig_2}c, \ref{fig_2}d, \ref{fig_3}c, \ref{fig_3}d, \ref{fig_4}c, \ref{fig_4}d, \ref{fig_5}c, and \ref{fig_5}d): in the MHD regime, rapid isotropization prevents energy from being stored in pressure anisotropy channels, allowing more energy to be funneled into Alfv\'{e}nic modes. The enhanced Alfv\'{e}nic activity leads to stronger bending of magnetic field lines, more efficient magnetic reconnection, and the formation of larger magnetic islands (plasmoids). These localized structures generate sharp spatial gradients in $B_y$, which manifest as the extended wings in the PDF and indicate higher intermittency. In contrast, in the CGL limit, a fraction of the energy is accommodated in parallel and perpendicular pressure channels, reducing the energy available to drive reconnection. As a result, fewer magnetic islands form, the fluctuations in $B_y$ are smaller, and the PDFs are correspondingly narrower. These results indicate that the intermittency of the transverse magnetic field is governed by the interplay between pressure anisotropy and magnetic reconnection, with enhanced small-scale fluctuations observed in the MHD regime. Although intermittency is inherently more pronounced and meaningful in fully three-dimensional systems, these two-dimensional simulations provide a clear preliminary indication of how the degree of intermittency in the transverse magnetic field is modulated by pressure anisotropy and magnetic reconnection, and they hint at the rich dynamics that could be explored in future three-dimensional studies.

%%%%%%%% Section:6 Conclusions %%%%%%%%

\section{Conclusions \& Future Directions} \label{sec:conclusion}
\subsection{Conclusions}
The physics of dilute plasmas has always indicated that a such plasmas might not retain an isotropic pressure. Numerous lines of data from magnetospheric measurements around planets \citep{Bale_2009} have indicated that the plasma in the solar wind is susceptible to anisotropies leading to plasma depletion layers on the bow shock side of the magnetosphere. \citet{Cairns_2017} have also made a case for such anisotropies being present at the heliopause where the VLISM (Very Local Interstellar Medium) abuts the inner heliosheath. \citet{Ma_2025}, and other works cited therein, shown the importance of KH instabilities in the heliosheath. Those authors find that depending on the location in the heliosheath, one can have aligned as well as anti-aligned configurations of the magnetic field. For that reason, we have also simulated anti-aligned as well as aligned cases for the KH instability in this paper. A stand-alone study of the KH instability within the context of the CGL equations is very important, especially if it is to show significant differences from the KH instability for MHD. Due to previous numerical limitations, such a study had indeed been out of reach. However, in \citet{Bhoriya_Balsara_Florinski_Kumar_2024} we showed that the simulations of the CGL \citep{Chew_Goldberger_Low_1956} equations has indeed become feasible. The relaxation time plays a very important role in understanding the physics that develops, which is why we have explored a range of relaxation times spanning several eddy turnover times to a vanishingly small fraction of an eddy turnover time (i.e. the MHD limit).

After describing our set-up in Section \ref{sec:model}, we have outlined a linear stability analysis of the KH instability in Section \ref{sec:linertheory}. The results of the match-up between the linearised growth rates and numerical growth rates are presented in Section \ref{subsec:growth rate}. A good concordance is found between the linearised and numerical growth rates, which gives us a good verification of the numerical capabilities.

We have also visually analyzed the data in Section \ref{sec:results}, which is then followed up by quantitative analysis of the same in Section \ref{sec:diagnostics}. In all instances, we find that the KH instability shows maximal growth in the MHD limit, and that growth rate is strongly suppressed when the relaxation time is some non-negligible fraction of the turn over time. Simulations in the CGL limit show rapid and prominent development of pressure anisotropies in the flow, a trend that is absent in the MHD limit. Initial growth favors anisotropy ratios that are larger than unity, with anisotropy ratios less than unity being favored in the late stages of instability development. This is also documented by quantitative analysis of the data that includes bulk analysis and pressure plots of the numerical data.

For the anti-aligned cases, the current densities also show more intensity in the MHD limit than in the CGL limit, resulting in the formation of larger magnetic islands. While magnetic island formation is suppressed for the aligned case, even there we see that the KH instability in the MHD limit is more vigorous than it is in the CGL limit. The physical explanation for that is as follows: In the CGL limit, some portion of the non-linear turbulent energy can go towards the formation of different regions with strong pressure anisotropies. However, this shuttling of energy in the pressure modes deprives the magnetic fields of energy that is needed for their bending. As a result, our overarching finding is that in both the aligned and anti-aligned magnetic field configurations the CGL simulation show reduced magnetic activity compared to the simulations in the MHD limit. It is also possible that the predominance of perpendicular pressure in the CGL siluations reduces the rate of plasma inflow into the current sheet that further inhibits reconnection.

While our 2D simulations are not best-positioned to talk about intermittency effects, we also find that certain diagnostics of intermittency are suppressed in the CGL limit vis a vis the MHD limit. 
The modification of the KH instability due to pressure anisotropy within the CGL framework has been previously established by \citet{Kruznetsov_1995}. It is crucial to underscore that these findings, similar to the current investigation, are constrained by the recognized limits of the CGL formalism. In particular, CGL fails to accurately replicate the mirror instability, with the mirror-instability threshold deviating from the kinetic result by approximately a factor of 6 \citep{Snyder_1997}. This constraint should be considered while analyzing results related to mirror-instability effects, if the system displays substantial anisotropy ratios.

\subsection{Future Directions}
In space plasmas, the contrast between the CGL and MHD limits has important implications for particle acceleration. In the MHD regime, enhanced reconnection and the formation of magnetic islands create intermittent structures that provide efficient sites for particle energization. In contrast, the CGL simulations show stabilized channel flows with suppressed reconnection, reducing the availability of such sites and thereby limiting particle acceleration. This highlights the role of pressure anisotropy as a regulator of both intermittency and particle energization in collisionless space plasmas.

These results suggest that incorporating CGL-based descriptions into studies of heliosheath dynamics may provide new perspectives on the role of pressure anisotropy in shaping reconnection and intermittency. Such an approach could offer useful guidance for future efforts to interpret Voyager 1 \& Voyager 2 observations and improve our understanding of particle acceleration and energy transfer processes in the outer heliosphere. The CGL-based calculations may also have important implications for understanding magnetospheric dynamics of planets, where pressure anisotropy can significantly influence reconnection, turbulence, and particle acceleration.

Beyond their relevance to space physics, these studies also carry important implications for broader astrophysical contexts, where pressure anisotropy plays a critical role in shaping the plasma dynamics.
For example, in accretion disk plasmas, the low collisionality renders the system susceptible to pressure anisotropy. In such plasmas, the magnetorotational instability (MRI) is considered the dominant mechanism driving turbulent viscosity, which in turn facilitates accretion flows. If magnetic activity is suppressed in the CGL limit relative to the MHD limit, as suggested by our present simualtion results, one may expect that the effective $\alpha$ parameter characterizing angular momentum transport in accretion disks would be reduced in the CGL regime. Hence, these findings underscore the broader relevance of pressure-anisotropy effects, highlighting their potential impact not only in heliospheric and planetary contexts but also in astrophysical systems such as accretion disks and interstellar turbulence, where they can modulate magnetic activity, turbulence, and energy transport.

We have seen that many diagnostics vary with increasing amounts of relaxation time. However, some diagnostics, like those shown in Fig. \ref{fig_6} or Fig. \ref{fig_7}, are not very strongly dependent on the value of the relaxation time. But it does matter that some reasonable relaxation time is included. This reminds us of the role of cooling times in astrophysical simulations. The important thing is that cooling is included, but the details of the cooling function are not very relevant. In an analogous way, the details of the relaxation time do not matter much when we are considering certain diagnostics. The big differentiator is that some relaxation time should be included in order to enable the plasma to have a range of anisotropic pressures. The authors realize that they are just mentioning their intuitive anticipation here. It will require many more simulations before this can be verified. It is also important to note that, in the present study, we have primarily focused on the effect of the CGL relaxation time on the evolution of the two-dimensional KH instability, and have therefore considered only longitudinal perturbations. While it would be of considerable interest to investigate the evolution of the KH instability in fully three-dimensional simulations including oblique perturbations, this lies beyond the scope of the current work. We intend to address this problem in future communications.

We conclude that CGL simulations open several intriguing avenues for future research, as they fundamentally modify plasma dynamics in ways that impact both space physics and astrophysical systems. These effects, often captured only within the MHD framework, underscore the importance of accounting for pressure anisotropy when understanding plasma behavior across diverse environments.

%\begin{acknowledgments}
\section{Acknowledgments}
All the computer simulations were performed at the Center for Research Computing cluster at UND. DSB acknowledges funding from NASA grant 4500005314 and NSF grants NSF-AST-2434532 and NSF-OAC-2514038. The authors were supported by NASA grant 18-DRIVE18\_2-0029, Our Heliospheric Shield, 80NSSC22M0164.
%\end{acknowledgments}

\bibliography{manuscript}{}
\bibliographystyle{aasjournal}

\end{document}